\documentclass[pra,twocolumn,superscriptaddress]{revtex4}
\usepackage{amsfonts}
\usepackage{amsmath}
\usepackage{amsthm}
\usepackage{amscd}
\usepackage{amssymb}
\usepackage{subfigure}
\usepackage{amsxtra}
\usepackage{bm}           
\usepackage{bbm}
\usepackage{exscale}
\usepackage{epic,rotating}
\usepackage{graphics,color,dsfont}
\usepackage{latexsym}
\usepackage{enumerate}
\usepackage{dcolumn}      
\usepackage{pstricks,pst-grad,fancybox,graphics}
\usepackage[dvips]{epsfig}
\usepackage[crop=off]{auto-pst-pdf} 

\def\bi#1\ei {\begin{itemize}#1\end{itemize}}
\def\bn#1\en {\begin{enumerate}#1\end{enumerate}}
\def\bea#1\eea {\begin{align}#1\end{align}}
\def\bean#1\eean {\begin{align*}#1\end{align*}}
\def\ben#1\een {\begin{equation*}#1\end{equation*}}
\def\be#1\ee {\begin{equation}#1\end{equation}}
\def\bes#1\ees {\begin{equation}\begin{split}#1\end{split}\end{equation}}
\def\bear#1\eear {\begin{eqnarray}#1\end{eqnarray}}
\def\bear#1\eear {\begin{eqnarray*}#1\end{eqnarray*}}

\newcommand{\bra}[1]{\ensuremath{\langle#1|}}
\newcommand{\ket}[1]{\ensuremath{|#1\rangle}}

\newcommand{\mean}[1]{\ensuremath{\langle #1 \rangle}}

\newcommand{\ex}[1]{\ensuremath{\mbox{e}^{#1}}}
\newcommand{\eins}{\ensuremath{\mathbbm 1}}

\newcommand{\tr}[1]{\ensuremath{\mbox{Tr}\left\{ #1 \right\}}}

\newcommand{\RRe}[1]{\textnormal{Re}\left(#1\right)}
\newcommand{\IIm}[1]{\textnormal{Im}\left(#1\right)}

\begin{document}

\title{Non-classicality tests and entanglement witnesses for macroscopic mechanical superposition states}

\author{Oleg Gittsovich}
\affiliation{Atominstitut - Institute of Atomic and Subatomic Physics, Vienna University of Technology, Stadionallee 2, A-1020 Vienna, Austria}

\author{Tobias Moroder}
\affiliation{Naturwissenschaftlich-Technische Fakult\"at, Universit\"at Siegen, Walter-Flex-Stra\ss e 3, D-57068 Siegen, Germany}

\author{Ali Asadian}
\affiliation{Atominstitut - Institute of Atomic and Subatomic Physics, Vienna University of Technology, Stadionallee 2, A-1020 Vienna, Austria}
\affiliation{Naturwissenschaftlich-Technische Fakult\"at, Universit\"at Siegen, Walter-Flex-Stra\ss e 3, D-57068 Siegen, Germany}

\author{Otfried G\"uhne}
\affiliation{Naturwissenschaftlich-Technische Fakult\"at, Universit\"at Siegen, Walter-Flex-Stra\ss e 3, D-57068 Siegen, Germany}

\author{Peter Rabl}
\affiliation{Atominstitut - Institute of Atomic and Subatomic Physics, Vienna University of Technology, Stadionallee 2, A-1020 Vienna, Austria}

\begin{abstract}
We describe a set of measurement protocols for performing non-classicality tests and the verification of entangled superposition states of macroscopic continuous variable systems, such as nanomechanical resonators. Following earlier works, we first consider a setup where a two-level system is used to indirectly probe the motion of the mechanical system via Ramsey measurements and discuss the application of this methods for detecting non-classical mechanical states. We then show that the generalization of this techniques to multiple resonator modes allows the conditioned preparation and the detection of entangled mechanical superposition states. The proposed measurement protocols  can be implemented in various qubit-resonator systems that are currently under experimental investigation and find applications in future tests of quantum mechanics at a macroscopic scale.
\end{abstract}

\maketitle

\section{Introduction}
Quantum superpositions of massive particles and non-classical correlations associated with entangled states are two of the most fascinating aspects that distinguish quantum mechanics from preceding classical theories.
While by now these concepts are well established and experimentally verified with high precision with photons~\cite{Aspect1999,Haroche2013,Kirchmair2013}, atoms~\cite{Kasevich1991,WinelandRMP2013,Monz2011,Robens2014} or molecules~\cite{Arndt2014}, there is still a strong interest in whether or not the laws of quantum mechanics are equally valid on a macroscopic scale~\cite{OxfordQuestions}. Various collapse models~\cite{GPW,Diosi,Penrose,GhirardiPRA1990,Bassi2013} predict a breakdown of the superposition principle at a certain mass and length scale, but so far testing these predictions has been beyond current experimental capabilities. Recently, micro- and nanomechanical resonators with masses in the picogram regime have been cooled close to the quantum ground state~\cite{OConnel2010,Teufel2011,Chan2011}, entangled with microwave photons~\cite{Palomaki2013}, and first steps towards a coherent coupling between mechanical systems and spin~\cite{ArcizetNatPhys2011,KolkowitzScience2012,Ovartchaiyapong2014,TeissierPRL2014} or charge based qubits~\cite{OConnel2010,LaHayeNature2009,PirkkalainenNature2013} have been implemented. These achievements  show that experiments with opto- and nanomechanical systems~\cite{Poot2012,Aspelmeyer2013} offer a promising route towards systematic tests of quantum mechanics with truly massive objects.

%

Due to the weak intrinsic nonlinearities of micro- and nanomechanical systems it is in general hard to prepare or probe nonclassical states in such systems directly. Thus, many of the initial  proposals for generating macroscopic superposition states considered the dispersive coupling of a mechanical resonator to a microscopic two level system (qubit)~\cite{ArmourPRL2002,TianPRB2005,MarshallPRL2003}. Provided that this coupling is sufficiently strong, it will evolve an initial qubit superposition state into an equal superposition of displaced resonator states and the survival of this superposition can be inferred from observing an initial loss and later revival of the qubit coherence.
%
%
In a recent proposal \cite{asadian_probing_2013} it has further been shown how the same type of coupling can be used to probe quantum superpositions of a mechanical resonator mode via Ramsey correlation measurements.
In the protocol of Ref. \cite{asadian_probing_2013} the non-classicality of the mechanical system is deduced directly from the violation of a Leggett-Garg-type inequality~\cite{LG85,Nori}. Thereby, such correlation measurements complement the less conclusive interference signatures mentioned above and provide a simple alternative to more involved schemes for implementing a complete tomography of the mechanical state~\cite{Rabl2004,Zhang2006,Singh2010,Tufarelli2011}.

In this paper we describe a generalization of the Ramsey measurement  technique for the detection of entanglement between two mechanical modes, in particular for verifying the entanglement between macroscopic superposition states (`Schr\"odinger cat states'), which will be most relevant in searches for hypothetical collapse mechanisms.
In the first part of this work we will first review the general idea of Ramsey measurements of mechanical motion and its connection to modular variables and the characteristic function. By employing the non-classicality criterion by Vogel \cite{vogel_nonclassical_2000} this relation can already be used to implement a simple measurement protocol that is capable of detecting many non-classical states of the nanomechanical oscillator without full state tomography~\cite{Agarwal2012}. In the second part we then apply a related strategy for constructing a witness for entangled superposition states. We first show that probing the characteristic function of two oscillators with this scheme in any two points of the space is not sufficient to detect entanglement, meaning that it is not possible to directly swap the entanglement from resonator modes onto an entangled two-qubit state in such a way. Therefore, in this work we identify a minimally extended set of measurements that can serve as a witness for entangled Schr\"odinger cat states and we provide particular examples of the states and of the measurement settings that are required to verify entanglement in those states.

The remainder of the paper is structured as follows. In Sec.~\ref{sec:RamseyMeasurements} we first summarize the basic idea of a Ramsey-type measurement of mechanical motion. In Sec.~\ref{sec:Nonclassicality} we illustrate the application of this method for detecting the non-classicality of a mechanical state in terms of two basic examples. Finally, Sec.~\ref{sec:Entanglement} contains the main results of this work and we discuss the protocols for generating and verifying the entanglement between mechanical superposition states. A summary of our findings and concluding remarks are given in Sec.~\ref{sec:Conclusions}.

\section{Ramsey measurements, modular variables and the characteristic function}\label{sec:RamseyMeasurements}

\begin{figure}[ht]
\includegraphics[width=.99\columnwidth]{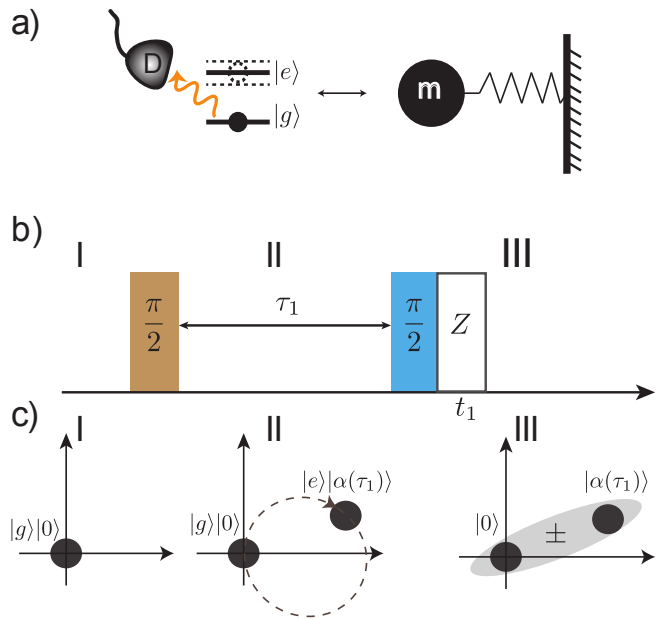}
\caption{(color online) a) Coupling between the two-level atom and the nanomechanical oscillator and its mechanical analog; b) Ramsey measurement consisting of three steps: I) an initial $\pi/2$ rotation, II) evolution under the Hamiltonian in Eq. (\ref{eq:hamilt}) for a time $\tau_1$, III) final $\pi/2$ rotation and a successive measurement of the atom population $Z$;
c) state evolution under the Ramsey sequence I - III as given by Eq. (\ref{eq:unitary}).}
\label{fig:setup}
\end{figure}

For the following discussion we consider a setup as schematically shown in Fig.~\ref{fig:setup} a), where a two level system (qubit) with ground state $|g\rangle$ and excited state $|e\rangle$ is coupled to the motion of a macroscopic mechanical resonator. We assume that the interaction between the two level system and the resonator is purely dispersive, i.e. the energy of the excited state is shifted proportional to the displacement of the resonator.  Then, in a frame rotating with the bare qubit splitting, $\omega_{eg}$, the system is described by the Hamiltonian $(\hbar=1)$
\be
H=\omega a^{\dagger}a + \lambda(a + a^{\dagger})|e\rangle\langle e|,
\label{eq:hamilt}
\ee
where $a, a^{\dagger}$ are the annihilation and the creation operators of the resonator mode, $\omega$ is the mechanical vibration frequency and $\lambda$ is the interaction strength. The type of coupling given in Eq.~\eqref{eq:hamilt} appears in various different scenarios where micro- and nanomechanical resonators are coupled to electronic spins~\cite{Rabl2009,ArcizetNatPhys2011,KolkowitzScience2012}, quantum dots~\cite{WilsonRaePRL2004,Yeo2014,Montinaro2014}, superconducting qubits~\cite{LaHayeNature2009,ArmourPRL2002,TianPRB2005},  or photons~\cite{MarshallPRL2003}.
For the following discussion the specific physical realization of Hamiltonian~\eqref{eq:hamilt} is not of immediate importance, and the reader is referred  to the above listed references for more details on possible implementations.
%

\subsection{Ramsey measurements}\label{ss:measscheme}
Hamiltonian~\eqref{eq:hamilt} describes a frequency shift of the excited qubit state which is proportional to the displacement of the mechanical resonator. This frequency shift can be detected via a Ramsey measurement performed on the qubit~\cite{SteinkePRA2011,BennettNJP2012}, which thereby serves as a readout device for the mechanical mode.
This method has already been used to detect, for example, the driven and thermal motion of mechanical systems in the classical regime~\cite{KolkowitzScience2012,Ovartchaiyapong2014,Qu2014}. Here we are interested in a full quantum mechanical description of this measurement.

Starting with the qubit initialized in state $|g\rangle$, the Ramsey sequence consist of four steps, which are depicted in Fig.~\ref{fig:setup} b) and c).  i) First, a fast $\pi/2$ rotation $R_{\frac{\pi}{2}}(\phi_0)$ prepares the qubit in state $R_{\frac{\pi}{2}}(\phi_0)|g\rangle = (|g\rangle + e^{i\phi_0}|e\rangle)/\sqrt{2}$. ii) The qubit-resonator system then evolves under the action of Hamiltonian~\eqref{eq:hamilt} for a time $\tau$.  The corresponding evolution operator $U(\tau)=e^{-iH\tau}$ can be written as
\begin{equation}\label{eq:U}
U(\tau)= \left[ \mathbbm{1}  \otimes |g\rangle\langle g| + e^{i  \phi_g}\mathcal{D}(\alpha)\otimes |e\rangle\langle e|\right] U_0(\tau),
\end{equation}
where $\mathcal{D}(\alpha)=e^{\alpha a^\dag -\alpha^* a}$ is the displacement operator, $U_0(\tau)= e^{-i\omega \tau a^\dag a}$ is the free resonator evolution and $\phi_g=\lambda^2/\omega^2(\omega\tau-\sin \omega\tau )$ is a geometric phase.  Eq.~\eqref{eq:U} represents a state dependent displacement of the resonator mode by an amount $\alpha=\lambda/\omega(e^{-i\omega \tau}-1)$, and evolves the initial qubit superposition into an equivalent superposition of displaced resonator states  as indicated in the middle panel of Fig.~\ref{fig:setup} c). iii) Finally, the qubit is rotated by another  $\pi/2$ pulse, $R_{\frac{\pi}{2}}(\phi_0=0)$, and iv) the state of the qubit (in the Z-basis) is detected.

In summary, starting at time $t=0$ with the qubit in $|g\rangle$ and the resonator mode in an arbitrary state $\rho_m$ the Ramsey measurement implements the combined unitary operation
\be
U_{\rm R}(\tau,\phi_0)=R_{\frac{\pi}{2}}(0)U(\tau) R_{\frac{\pi}{2}}(\phi_0),
\label{eq:unitary}
\ee
followed by a projective measurement of the qubit state. The whole sequence can thus be described as a generalized measurement~\cite{asadian_probing_2013} on the resonator mode, where the probability $p_+$ ($p_-$) for finding the qubit in the excited (ground) state is given by
\begin{equation}
p_\pm = \tr{E_\pm^\dag(\varphi,\tau) E_\pm(\varphi,\tau) \rho_m},
\label{eq:probqubit}
\end{equation}
and conditioned on the measurement outcome the resonator state is projected into one of the states
\begin{equation}
\rho_m^\pm = \frac{E_\pm(\varphi,\tau)\rho_m E_\pm^\dag(\varphi,\tau)}{p_\pm}. 
\label{eq:condstate}
\end{equation}
In Eq. (\ref{eq:probqubit}) and (\ref{eq:condstate})  $E_\pm(\varphi,\tau) =\frac{1}{2}\left[\mathbbm{1} \pm e^{i \varphi}\mathcal{D}(\alpha) \right]U_0(\tau)$
are Kraus operators satisfying $E_+^\dag E_++E_-^\dag E_-=\mathbbm{1}$,  and $\varphi=\phi_0 +\phi_g$ is the total phase. Eq.~\eqref{eq:condstate} shows that this technique can not only be used to probe mechanical motion, but also to prepare -- conditioned on the outcome -- a mechanical superposition state. In particular, when the resonator is initially prepared close to the ground state, $\rho_m=|0\rangle\langle 0|$, it is projected after the measurement into one of the two superposition states~\cite{TianPRB2005,SteinkePRA2011,BennettNJP2012,VacantiPRA2013,asadian_probing_2013}
\begin{equation}\label{eq:Superposition1}
 |\psi^\pm\rangle = \frac{|0\rangle \pm e^{i \varphi} |\alpha\rangle}{\sqrt{4p_\pm}}.
\end{equation}
Note that while for the static coupling given in Eq. (\ref{eq:hamilt}) $|\alpha| \leq 2\lambda/\omega$, the displacement amplitude can be resonantly enhanced by periodically flipping the qubit state during the interaction time $\tau$~\cite{TianPRB2005,BennettNJP2012,asadian_probing_2013}. Thus, in the following we will consider $\alpha$ as an adjustable parameter. In  practice the magnitude of the displacement will eventually be limited by the qubit coherence time $T_2$ and the mechanical rethermalization rate $\Gamma_m\simeq k_BT/(\hbar Q)$, where $T$ is the support temperature and $Q$ the mechanical quality factor~\cite{TianPRB2005,asadian_probing_2013}.

\subsection{Modular variables and the characteristic function}\label{ss:modvarcf}
For the following discussion it is convenient to re-express  Eq.~\eqref{eq:probqubit}  in terms of the average population difference $\langle Z\rangle=p_+-p_-$, which can then be written as
\begin{align}
\langle Z\rangle(\varphi,\alpha) =\tr{Q(\varphi,\alpha)\rho_m}.
\end{align}
Here
\begin{equation}\label{eq:Qdef}
Q(\varphi,\alpha)=\frac{1}{2}
\left(
\ex{i\varphi}\mathcal{D}(\alpha)+\ex{-i\varphi}\mathcal{D}^{\dagger}(\alpha)\right)
\end{equation}
is a modular operator~\cite{aharonovrohrlichBook}  that can also be expressed in terms of the position and the moment operators $\hat{x}=(a+a^\dag)/\sqrt{2}$ and $\hat{p}=i(a^\dag-a)/\sqrt{2}$ as
\begin{align}
Q(\varphi,\alpha)=\cos\left(\varphi + \sqrt{2}\IIm{\alpha}\hat{x} - \sqrt{2}\RRe{\alpha}\hat{p}\right).
\end{align}
Thus, for appropriately chosen $\varphi$ and $\alpha$, the measurement of $\langle Z\rangle$ directly probes expectation values $\langle \cos(|\alpha|Ê\hat x)\rangle$, $\langle \sin(|\alpha| \hat x)\rangle$, $\langle \cos(|\alpha| \hat p)\rangle$, etc. From Eq.~\eqref{eq:Qdef} it also follows immediately that the Ramsey scheme can be used to  measure the characteristic function $\chi (\alpha)= \tr{\mathcal{D}(\alpha) \rho_m}$ via the relation
\begin{align}
\chi(\alpha)=\langle Z(\varphi=0,\alpha)\rangle  + i  \langle Z(\varphi=-\pi/2,\alpha)\rangle.
\end{align}
The symmetrically ordered characteristic function $\chi (\alpha)$ is the Fourier transform of the Wigner function,
\begin{equation}\label{eq:Wignerfunction}
W(\xi)= \int\frac{d^2 \alpha}{\pi^2}\chi (\alpha)e^{\xi\alpha^{\ast}-\xi^{\ast}\alpha},
\end{equation}
and therefore the knowledge of $\chi(\alpha)$ for a sufficiently dense set of points $\alpha$ in phase space would allow a complete reconstruction of the mechanical state~\cite{Zhang2006,Singh2010,Tufarelli2011}.

This formal connection to the characteristic function will also hold for multi-mode systems. In this work we are primarily interested in the case, where the Ramsey sequences is simultaneously carried out with two qubits, each coupled to one resonator mode.  Denoting by $\varphi_1$ and $\varphi_2$ the adjustable phases and by $\alpha$ and $\beta$ the displacement amplitudes in the two measurements, respectively,  the combined outcome is
\begin{equation}
\langle Z_1 Z_2\rangle = \tr{Q(\varphi_1,\alpha) \otimes Q(\varphi_2, \beta)\rho_{m_1m_2}},
\end{equation}
where $\rho_{m_1m_2}$ is the total density operator of the two mechanical modes. These averages can again be used to extract the
two-mode characteristic function $\chi(\alpha,\beta)=\tr{\mathcal{D}(\alpha)\mathcal{D}(\beta) \rho_{m_1m_2}}$ by using the relations
\bea
\RRe{\chi(\alpha,\beta)}=&\mean{Q(0,\alpha)\otimes Q(0,\beta)}\nonumber\\
&-\mean{Q(-\pi/2,\alpha)\otimes Q(-\pi/2,\beta)},\\
\IIm{\chi(\alpha,\beta)}=&\mean{Q(0,\alpha)\otimes Q(-\pi/2,\beta)}\nonumber\\
&+\mean{Q(-\pi/2,\alpha)\otimes Q(0,\beta)}.\label{eq:Chi2}
\eea
In principle this scheme allows the reconstruction of the full two-mode Wigner-function (full tomography), for which a number of entanglement criteria exist.
However,  in practice the full tomography requires a lot (in fact infinitely many) of measurements and enough statistic for the state reconstruction. In the following we will discuss simpler witnesses for non-classicality and entanglement, which are based on the relations described in this section, but require the measurement of only a few expectation values.

\section{Non-classicality tests for nanomechanical oscillators}\label{sec:Nonclassicality}
Before addressing the question of entanglement verification in Sec.~\ref{sec:Entanglement}, we first describe in this section the application of the Ramsey method for testing the non-classicality of a single resonator mode.
%
In quantum optics one usually speaks of a \emph{non-classical state}~\cite{Titulaer1965,Mandel1986} if the corresponding $P$-function  defined via
\begin{equation}
\rho= \int d^2\xi  \, P(\xi) |\xi\rangle\langle \xi|,
\end{equation}
does not represent a proper (non-negative and normalized) probability distribution in phase space. In analogy to the Wigner function, the $P$-function is given by the Fourier-transform of the normally-ordered characteristic function $\chi_N(\alpha)=\mean{:\mathcal{D}(\alpha):}=  e^{\frac{1}{2}|\alpha|^2} \mean{D(\alpha)}$, where the symbol $:\::$ stands for normal ordering of operator products.  It follows that for a \emph{classical} state
\begin{align}
|\mean{:\mathcal{D}(\alpha):}|= & \left|\int d^2\xi \, P(\xi) e^{-(\xi\alpha^{\ast}-\xi^{\ast}\alpha)}\right|\\
\leq & \int d^2\xi \, P(\xi) \left|e^{-2i\IIm{\xi\alpha^{\ast}}}\right|\leq  1.
\end{align}
In other words, this relation means that for a classical state the corresponding characteristic function is bounded by
\begin{equation}\label{eq:NC1}
|\chi(\alpha)|=|\mean{\mathcal{D}(\alpha)}|\leq e^{-|\alpha|^2/2},
\end{equation}
i.e. it decays faster than the characteristic function of the ground state. The violation of this inequality is a sufficient, but not necessary~\cite{Diosi2000} condition for non-classical states.

\begin{figure}
\includegraphics[width=.45\textwidth]{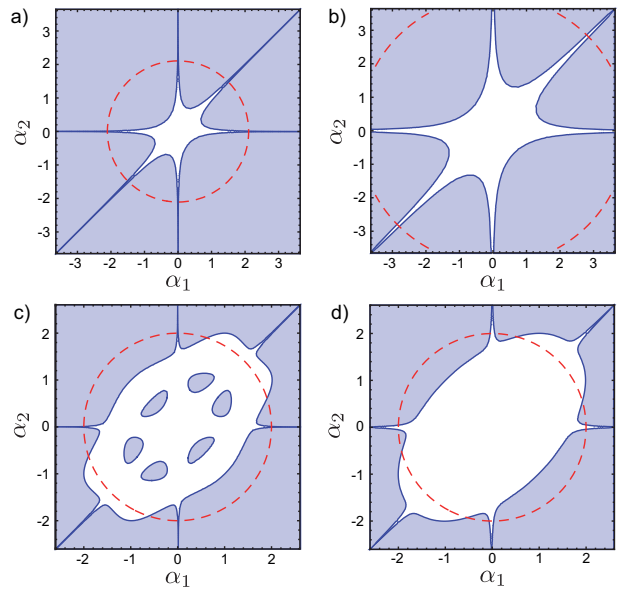}
\caption{(color online). Non-classicality detection for the state $\rho_n=(1-p)\ket{n}\bra{n}+p\ket{0}\bra{0}$. The shaded parts indicate the regions where the criterion given by Eq.~\eqref{eq:NC2} is violated. The upper two panels show the result for $n=1$ and the lower two panels for $n=2$. The values for $p$ are $p=0.1$ in a) and c) and $p=0.75$ in b) and d) and $\alpha_i \in \mathbbm{R}$. The red dashed circle indicates the value of $|\alpha|$ that is required to detect non-classicality using the criterion given in Eq.~\eqref{eq:NC1}.}
\label{fig:detection_regions}
\end{figure}

As shown in Ref.~\cite{vogel_nonclassical_2000,richter_nonclassicality_2002}, this bound can be generalized and further improved via the classical Bochner-Khinchin theorem. Applied to this scenario, it states that for any classical state $\rho$ and for any set of test points $\{ \alpha_i\}$, the matrix $M_{ij} = \langle  :\!D(\alpha_i - \alpha_j)\!: \rangle$ is positive semidefinite. Thus a violation immediately certifies non-classicality of the underlying system.
%
Depending on the number of test points one gets different detection strengths. If one only takes $\{\alpha_0=0, \alpha_1\}$, then positivity of the resulting $2 \times 2$ matrix is equivalent to the bound given in Eq.~\eqref{eq:NC1} above, which is also the criterion of Refs.~\cite{vogel_nonclassical_2000,Agarwal2012}.
But if we use more points, like  $\{ \alpha_0=0, \alpha_1, \alpha_2\}$ with $\alpha_1 \not = \alpha_2\neq 0$, and apply the quantum (non-commutative) version of the Bochner-Khinchin theorem (see e.g. Chapter 5.4 in Ref. \cite{HolevoBook11}) then the  positivity requirement of
\begin{equation}
\label{eq:NC2}
\left[\!\begin{array}{ccc}
1 &\! \mean{:\mathcal{D}(-\alpha_1):} & \mean{:\mathcal{D}(-\alpha_2):}\\
\mean{:\mathcal{D}(\alpha_1):} &\! 1 & \mean{:\mathcal{D}(\alpha_1-\alpha_2):}\\
\mean{:\mathcal{D}(\alpha_2):} &\! \mean{:\mathcal{D}(\alpha_2-\alpha_1):} & 1
\end{array}\!\right]\! \geq 0
\end{equation}
gives a strictly stronger condition~\cite{richter_nonclassicality_2002}. This can be seen for instance by evaluating the criteria for the Fock states, where $\mean{:\mathcal{D}(\alpha):}_{\ket{n}}=L_n(|\alpha|^2)$ is given by the Laguerre polynomial of order $n$. The states $\rho_n=(1-p)\ket{n}\bra{n}+p\ket{0}\bra{0}$,
are non-classical for all values of $n\geq1$ and $p<1$, and for $n=1$ the condition given by Eq.~\eqref{eq:NC1} certifies non-classicality if $|\alpha_1| > \sqrt{2/(1-p)}$. For $n=2$, Eq.~\eqref{eq:NC1} certifies non-classicality for $|\alpha_1| > 2$ for all values $p<1$.
For the same states we plot in Fig.~\ref{fig:detection_regions} the non-classical region determined by the positivity criterion  in Eq.~\eqref{eq:NC2}. Note that for $n=1$ this criterion is in principle violated for all $\alpha_i$ and $p<1$, but for $|\alpha_i|\rightarrow 0$ the violation becomes very small and cannot be detected in realistic experiments. Therefore, the plots in Fig.~\ref{fig:detection_regions} show the regions where $\det(M)\leq -0.01$. We see that in many (but not all) cases, Eq.~\eqref{eq:NC2} allows us to identify regions where non-classicality can be certified with significantly smaller values for $|\alpha_i|$. For practical implementations of such test this can be very important:  By using more settings the required displacement amplitudes $\alpha_i$ and therefore the required qubit-resonator coupling strength $\lambda$ can be significantly lower than for non-classicality tests based on inequality~\eqref{eq:NC1} alone.

Below we are mainly interested in superposition states -- so-called Schr\"odinger cat states -- of the form
\begin{equation}\label{eq:SCState}
\ket{\psi_+}=\dfrac{1}{\sqrt{4p_+}}(\ket{0}+e^{i\theta} \ket{\xi_0}),
\end{equation}
where $p_+=(1+\cos(\theta)e^{-|\xi_0|^2/2})/2$. The normally-ordered characteristic function of this state is
\begin{align}
\chi_N(\alpha) =\frac{ 1+ e^{i2{\rm Im}( \alpha \xi_0^*)} + \left( e^{i\theta} e^{-\alpha^*\xi_0} + e^{-i\theta} e^{\alpha\xi_0^*}\right)e^{-\frac{|\xi_0|^2}{2}}
}{4p_+},
\end{align}
and it significantly exceeds the classical bound of Eq.~\eqref{eq:NC1} for values $|\alpha|\gtrsim |\xi_0|/2$ (see also Ref.~\cite{Agarwal2012}). If the resonator mode is weakly coupled to an environment, the superposition will decoher. In a frame rotating with the mechanical frequency $\omega$, the characteristic function will then evolve over time as
\begin{equation}
\label{eq:onerescat}
\chi_N(\alpha,t)= e^{-N_{\rm th}(1-e^{-\gamma t})|\alpha|^2}Ê\chi_N(\alpha e^{-\gamma t/2}),ÊÊ
\end{equation}
where $N_{\rm th}=1/(e^{\hbar\omega/k_BT}-1)$ is the thermal occupation number for an environment temperature $T$ and $\gamma$ is the mechanical damping rate.

\begin{figure}
\centering
\includegraphics[width=0.98\linewidth]{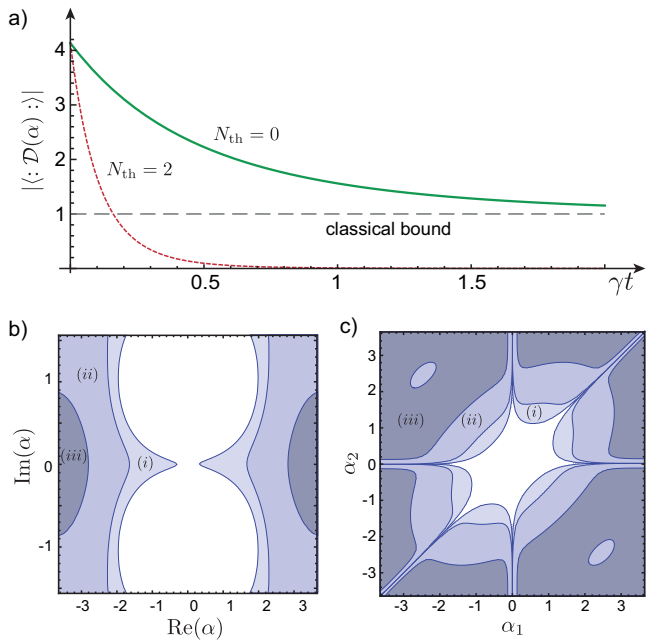}
\caption{(color online). Non-classicality of a Schr\"odinger cate state as defined in Eq.~\eqref{eq:SCState}. a) Time dependence of $\chi_N(\alpha,t)=\langle: \mathcal{D}(\alpha):\rangle(t)$ in the presence of decoherence and for $\alpha=\xi_0=2$, $\theta=0$. b) and c) Regions of non-classicality that can be detected using the criteria \eqref{eq:NC1} and \eqref{eq:NC2}, respectively.  In both plots the different shadings represent the non-classical regions evaluated at times (i) $\gamma t=0$, (ii) $\gamma t=0.02$, (iii) $\gamma t=0.04$ and assuming $N_{\rm th}=10$.}
\label{fig:NCCat}
\end{figure}

In Fig.~\ref{fig:NCCat} we plot the non-classical regions for a pure and partially decohered cat state for the example $\xi_0=3$. Fig.~\ref{fig:NCCat} a) shows the typical time dependence of the the non-classicality criterion Eq.~\eqref{eq:NC1}, evaluated for $\alpha=\xi_0$. Interestingly, the non-classicality of Schr\"odinger cat state coupled to a zero temperature bath never vanishes and approaches the classical bound asymptotically on a time-scale $\gamma^{-1}$. In contrast, for finite $N_{\rm th}>0$ the non-classical signatures are lost quickly on a timescale $(\gamma N_{\rm th}|\alpha|^2)^{-1}$. Similar results have been found in Ref.~\cite{Paavola2011}. Fig.~\ref{fig:NCCat} b) and c) compare the non-classicality criteria from Eq.~\eqref{eq:NC1} and Eq.~\eqref{eq:NC2} for $\xi_0=2$ and different stages of decoherence. In Fig.~\ref{fig:NCCat} c) we see similar patterns  as for the $n=2$ number state, but given that the maximal amplitude in Eq.~\eqref{eq:NC2} is $|\alpha_1-\alpha_2|$ the benefits compare to the simple criterion are not as big for the cat state.

\section{Entanglement witnesses}\label{sec:Entanglement}
Let us now consider the application of the Ramsey technique for the detection of entanglement between two mechanical modes. More precisely, in this work we are interested in entangled superposition states of the form
\begin{equation}\label{eq:EntangledCat}
\ket{\psi_+}=\frac{1}{\sqrt{2+2\ex{-4|\xi_0|^2}}}(\ket{\xi_0,\xi_0}+\ket{-\xi_0,-\xi_0}),
\end{equation}
which, apart from an overall shift in phase space, are the two-partite-entangled generalization of the Schr\"odigner cat state in Eq.~\eqref{eq:SCState}. This state combines the quantum mechanical principles of superposition and entanglement and it would thus be interesting to see in future studies how these two properties behave as $|\xi_0|$ or the mass of the mechanical system is increased.

\subsection{Preparation of two-mode entangled states between two nanomechanical resonators}\label{ss:prep}
Although in this work we are primarily interested in the entanglement detection scheme, we first briefly outline, how a state of the form given in Eq.~\eqref{eq:EntangledCat} can be prepared in probabilistic way by making use of the techniques described in Sec.~\ref{sec:RamseyMeasurements}.  Related schemes based on different type of resonator-qubit interactions have been discussed in the context of cavity QED~\cite{Davidovich1993}.

For the following protocol we consider the extension of Hamiltonian~\eqref{eq:hamilt} to two mechanical resonators, each coupled to its own qubit. We start off by preparing the mechanical resonators modes in the same state $|\psi\rangle$, which could be the ground state or a coherent state, and the two qubits in the Bell state $\ket{\Phi_+}=(\ket{gg}+e^{i\Theta}\ket{ee})/\sqrt{2}$. If the two mechanical modes represent, for example, two vibrational modes of a single cantilever, the qubits could be coupled directly to prepare such a state. If the mechanical systems are far apart, the entangled qubit states can be mediated via photons using standard procedures discussed for implementing quantum communication protocols. In a next step we apply the Ramsey sequence, which implements the unitary operation $U_{\rm R}$ given in Eq.~\eqref{eq:unitary} to each subsystem, such that the state of full system becomes
\begin{widetext}
\bea
U_R(\tau,\varphi)^{\otimes 2}\ket{\psi,\psi}\ket{\Phi_+}
&=\frac{1}{\sqrt{2}}\left[E_-(\tau,\varphi)\otimes E_-(\tau,\varphi)+ \ex{i(\Theta-2\phi_0)} E_+(\tau,\varphi)\otimes E_+(\tau,\varphi)\right]\ket{\psi,\psi}\ket{gg}\nonumber\\
&+\frac{1}{\sqrt{2}}\left[E_-(\tau,\varphi)\otimes E_+(\tau,\varphi)+ \ex{i(\Theta-2\phi_0)} E_+(\tau,\varphi)\otimes E_-(\tau,\varphi)\right]\ket{\psi,\psi}\ket{ge}\nonumber\\
&+\frac{1}{\sqrt{2}}\left[E_+(\tau,\varphi)\otimes E_-(\tau,\varphi)+ \ex{i(\Theta-2\phi_0)} E_-(\tau,\varphi)\otimes E_+(\tau,\varphi)\right]\ket{\psi,\psi}\ket{eg}\nonumber\\
&+\frac{1}{\sqrt{2}}\left[E_+(\tau,\varphi)\otimes E_+(\tau,\varphi)+ \ex{i(\Theta-2\phi_0)} E_-(\tau,\varphi)\otimes
E_-(\tau,\varphi)\right]\ket{\psi,\psi}\ket{ee}\label{eq:stateprep}.
\eea
\end{widetext}
After this interaction we measure the state of both qubits, which projects the resonator modes into one of the superposition state corresponding to the four lines in Eq.~\eqref{eq:stateprep}.
%
%
%
For example, for $\Theta=2\phi_0$ and if we find both qubits in state $|g\rangle$, which occurs with probability $p_{--}=[1+\RRe{\ex{2i\varphi}\langle\psi|\mathcal{D}(\alpha)|\psi\rangle}]/4$, the resonator modes are projected into the entangled state
\be
\ket{\psi}=\frac{1}{2\sqrt{p_{--}}}\left[\eins\otimes\eins+\mathcal{D}(\alpha)\otimes \mathcal{D}(\alpha)\right]\ket{\psi,\psi}.
\label{eq:theentst}
\ee
The outcome where one qubit is in state $|g\rangle$ and the other one in state $|e\rangle$ would produce the same state, but with a relative minus sign.
Alternatively, if we repeat the protocol with an initial qubit state $$\ket{\Psi_-}=\frac{1}{\sqrt{2}}(\ket{ge}-\ex{2i\phi_0}\ket{eg}),$$ the resulting resonator state for the two different measurement outcomes is
\be
\ket{\psi'}=\frac{1}{2\sqrt{p_{--}}}(\eins\otimes \mathcal{D}(\alpha) \mp \mathcal{D}(\alpha)\otimes\eins) \ket{\psi,\psi}.
\ee
Therefore, for $\alpha=\xi_0$ and by preparing both resonator modes initially in the ground state $|\psi\rangle=|0\rangle$, these protocols allows us to prepare a Bell basis of entangled cat states $\sim (\ket{0,0}\pm\ket{\xi_0,\xi_0})$, $\sim (\ket{0,\xi_0}\pm\ket{\xi_0,0})$. Similar, for $\alpha=2\xi_0$ and $|\psi\rangle=|-\xi_0\rangle$, we obtain the symmetric form of these states as given in Eq.~\eqref{eq:EntangledCat}. By using other initial resonator states, also  more general types of entangled states can be prepared in this way.




\subsection{No-Go result for entanglement swapping}\label{ss:nogo}
In the previous preparation protocol the entanglement between  the qubit states is swapped onto the resonator modes, conditioned on the outcome of the  final qubit measurement. This would suggest to use the same interaction to reverse the process and swap the entanglement back from the mechanical modes onto the otherwise decoupled qubits for verification. However,  it turns out that the asymmetry in this protocol, namely that we can only measure the state of the qubits, prevents such a scheme.

To show that the evolution generated by Hamiltonian~\eqref{eq:hamilt} does not allow to swap entanglement between the resonator modes and the qubits in a deterministic way, we consider the evolution operator $U(\tau)$ given in Eq.~\eqref{eq:U}, but without the initial and the final $\pi/2$ pulses, since the local unitary rotations do not affect the entanglement in the system. Without loss of generality we can also omit the phase $\phi_g$ and the free resonator evolution $U_0(\tau)$ in the following discussion. Then, the resulting total system evolution during the interaction time $\tau$ can be written as

\be
U_{\rm tot}(\tau)=\sum_{i,j=g,e}\ket{i,j}\bra{i,j}\otimes V_{ij},
\label{eq:genunitary}
\ee
where $V_{gg}= \mathbbm{1}\otimes \mathbbm{1}$, $V_{eg}= \mathcal{D}(\alpha)\otimes \mathbbm{1}$, $V_{ge}=\mathbbm{1}\otimes \mathcal{D}(\beta)$ and $V_{ee}= \mathcal{D}(\alpha) \otimes \mathcal{D}(\beta)$ are operators acting on the two mechanical modes.

Now let $\rho_{q_1q_2}\otimes \rho_{m_1m_2}$ be the total initial state of two nanomechanical oscillators and the two qubits. After the evolution in Eq. \eqref{eq:genunitary} the reduced state of two qubits can be expressed as
\begin{widetext}
\bea
\rho_{q_1q_2}^{\textnormal{out}}&=\Lambda[\rho_{q_1q_2}]={\rm Tr}_{m1m2}\left\{U_{\rm tot}(\tau)\left(\rho_{q_1q_2}\otimes \rho_{m_1m_2}\right)U^\dag_{\rm tot}(\tau)\right\}\nonumber\\
&=\sum_{ijkl=g,e}\ket{i,j}\bra{k,l}\bra{ij}\rho_{q_1q_2}\ket{kl}\tr{ V^{\dagger}_{lk}V_{ij} \rho_{m_1m_2}}=\rho_{q_1q_2}\odot M^T
\label{eq:chanact},
\eea
\end{widetext}
where $M$ is the $4\times 4$ positive-definite matrix of moments, which properties we discuss in greater detail in the next section, and $\odot$ is the Hadamard product (element-wise multiplication). Note that this mathematical structure for channels has previously been observed for interacting spin gases (see e.g. \cite{Walder-Hartmann_2005}).

Let us recall that the Hadamard product of two matrices is positive semi-definite if both matrices are positive semi-definite. Moreover,  for two matrixes $A$ and $B$, $(A\odot B)^{\Gamma}=A^{\Gamma}\odot B^{\Gamma}$, where $(\cdot)^{\Gamma}$ denotes the partial transpose with respect to the first subsystem, i.e. $[A^{\Gamma}]_{ij,kl}=A_{kj,il}$. This implies for the state after the action of the channel (\ref{eq:chanact})
\be
\left(\rho_{q_1q_2}^{\textnormal{out}}\right)^{\Gamma}=\rho_{q_1q_2}^{\Gamma}\odot (M^T)^{\Gamma}.
\label{eq:mom1}
\ee
It is not difficult to see that in this example the partial transposition of the matrix of moments $(M^T)^{\Gamma}$ corresponds to the sign flip of the displacement in the displacement operator which acts on the first oscillator. This leaves its eigenvalues unchanged. Since the $4\times 4$ matrix $M$ as mentioned above is itself positive semi-definite, the entanglement of the qubit state $\rho_{q_1q_2}$ is unaffected by the channel in Eq. (\ref{eq:chanact}).

We conclude the qubits' initial state $\rho_{q_1q_2}$ doesn't change its entanglement properties no matter what the initial quantum state of two nanomechanical oscillators $\rho_{m_1m_2}$ is supplied to the protocol.
 Since according to the results of Sec.~\ref{ss:modvarcf} the reduced density matrix of two qubits $\rho_{q_1q_2}^{\textnormal{out}}$ in Eq. (\ref{eq:chanact}) contains the information of the two-resonator characteristic function $\chi(\alpha,\beta)$, this observation has an interesting implication: The knowledge of the value of the characteristic function in one single point $(\alpha,\beta)$ is not sufficient in order to verify entanglement between the two mechanical modes.
Therefore, our goal is now to identify an entanglement witness using an extended but still small set of measurement points $\{(\alpha_i,\beta_i)\}$.

\subsection{Matrices of moments: tool for entanglement verification}

The matrix of moments that appeared in Eq. (\ref{eq:mom1}) has been extensively used as a tool for revealing nonclassical properties of states in quantum mechanics, see e.g.
\cite{vogel_nonclassical_2000,richter_nonclassicality_2002,shchukin_inseparability_2005,miranowicz_inseparability_2009,richter_nonclassical_2007}.
The general form of a matrix of moments for an operator $X$ is given by:
\be
M_{ij}(X)=\tr{\mathcal{V}_i^{\dagger}\mathcal{V}_jX}=\mean{\mathcal{V}_i^{\dagger}\mathcal{V}_j}_X, \; i,j=1,...,\infty,
\ee
where the $\mathcal{V}_i'$s are some dense set of operators (acting on a single or on a multi-partite system), i.e. any other operator can be represented in terms of the $\mathcal{V}_i'$s. For example, displacement operators and the products of the type $(a^{\dagger})^ka^l$ for $k,l=0,1,...,\infty$ form such a set.
The implementation of the matrix of moments to entanglement verification in Refs. \cite{shchukin_inseparability_2005,miranowicz_inseparability_2009} requires however photon-number resolving detectors and although the non-classicality tests are formulated directly in terms of the characteristic function they are afflicted with the same disadvantage \cite{KieselVogel2011,KieselVogel12}.

Nevertheless two established facts \cite{shchukin_inseparability_2005,miranowicz_inseparability_2009} will be important for us in the following discussion:
i) $X$ is positive semi-definite then $M(X)$ is positive semi-definite,
and
ii) a state $\rho$ is separable ($\rho=\sum_kp_k\rho^A_k\otimes\rho^B_k$, $p_k\geq 0$, $\sum_kp_k=1$) iff the corresponding matrix of moments is also separable $M(\rho)=\sum_kp_kM(\rho^A_k)\otimes M(\rho^B_k)$, with bipartite operators $\mathcal{V}_i=\mathcal{V}_{i_1}\otimes\mathcal{V}_{i_2}$ and $M(\rho^{A/B})=\tr{\mathcal{V}_{i_{1/2}}^{\dagger}\mathcal{V}_{i_{1/2}}\rho^{A/B}}$ respectively. Formally one can write
\be
\rho=\sum_kp_k\rho^A_k\otimes\rho^B_k \Leftrightarrow M(\rho)=\sum_kp_kM(\rho^A_k)\otimes M(\rho^B_k).
\ee
Moreover one can show that if a finite dimensional sub-matrix $M_{\textnormal{sub}}(\rho_{AB})$ of the matrix of moments is non-separable, then $\rho_{AB}$ is necessarily entangled (see e.g. Ref. \cite{moroder10a}).

We can use this fact for entanglement verification in our context, where expectation values of displacement operators can be computed from the directly accessible experimental data.


\subsection{Entanglement witnesses for macroscopic superposition states}

As suggested by our no-go result in Section \ref{ss:nogo} we should probe the characteristic function of the nanomechanical oscillators in several points. In order to do that we have to expand the set of operators from a commutative set $\{\eins, \mathcal{D}(\alpha)\}$ to a non-commutative set on the each side. By taking a displacement operator in one more additional point we achieve this goal and arrive at the set $\{\eins, \mathcal{D}(\alpha_1), \mathcal{D}(\alpha_2)\}$, $\alpha_1\neq\alpha_2$, for each party. The corresponding expectation values can be written as a $9\times 9$ matrix of moments
\begin{widetext}
\begin{changemargin}{-3cm}{-3cm}
\be
M=\left\langle\left(\begin{array}{ccc}
\eins & \mathcal{D}^{\dagger}(\alpha_1) &\mathcal{D}^{\dagger}(\alpha_2)\\
\mathcal{D}(\alpha_1) & \eins & \mathcal{D}^{\dagger}(\alpha_2)\mathcal{D}(\alpha_1)\\
\mathcal{D}(\alpha_2) & \mathcal{D}^{\dagger}(\alpha_1)\mathcal{D}(\alpha_2) & \eins
\end{array}\right)\bigotimes
\left(\begin{array}{ccc}
\eins & \mathcal{D}^{\dagger}(\beta_1) & \mathcal{D}^{\dagger}(\beta_2)\\
\mathcal{D}(\beta_1) & \eins & \mathcal{D}^{\dagger}(\beta_2)\mathcal{D}(\beta_1)\\
\mathcal{D}(\beta_2) & \mathcal{D}^{\dagger}(\beta_1)\mathcal{D}(\beta_2) & \eins
\end{array}\right)
\right\rangle_{\varrho_{m_1m_2}}.
\label{eq:twooscmom}
\ee
\end{changemargin}
\end{widetext}
As we outlined in Section \ref{ss:modvarcf} each entry of this matrix can be obtained by performing local Ramsey sequences and measuring the population of the qubit afterwards. Note that in an experiment we would need to measure three different settings per side: those corresponding to $\mathcal{D}(\alpha_1)$, $\mathcal{D}(\alpha_2)$ and $\mathcal{D}^{\dagger}(\alpha_1)\mathcal{D}(\alpha_2)$, which corresponds to measuring 24 expectation values in total ~\footnote{The entry that corresponds to the product of two displacement operators can be obtained if we recall the canonical commutation relations: $\mathcal{D}(\alpha_2)\mathcal{D}(\alpha_1) = \ex{i\IIm{\alpha_2\alpha^{\ast}_1}}\mathcal{D}(\alpha_2+\alpha_1)$.}.
\begin{figure}
\centering
\subfigure[]{
\includegraphics[width=.99\columnwidth]{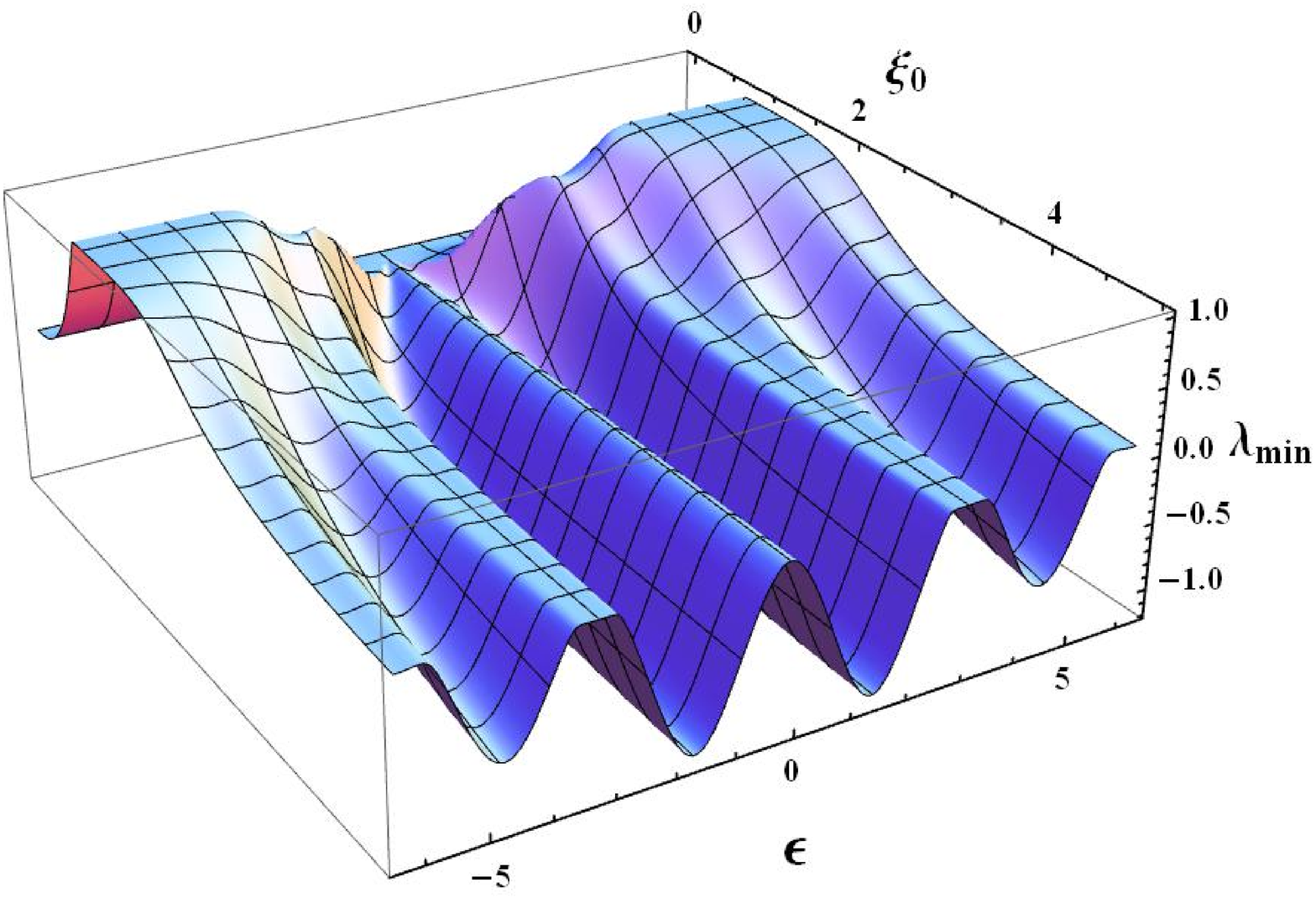}
\label{fig:minev3d}
}
\subfigure[]{
\includegraphics[width=.99\columnwidth]{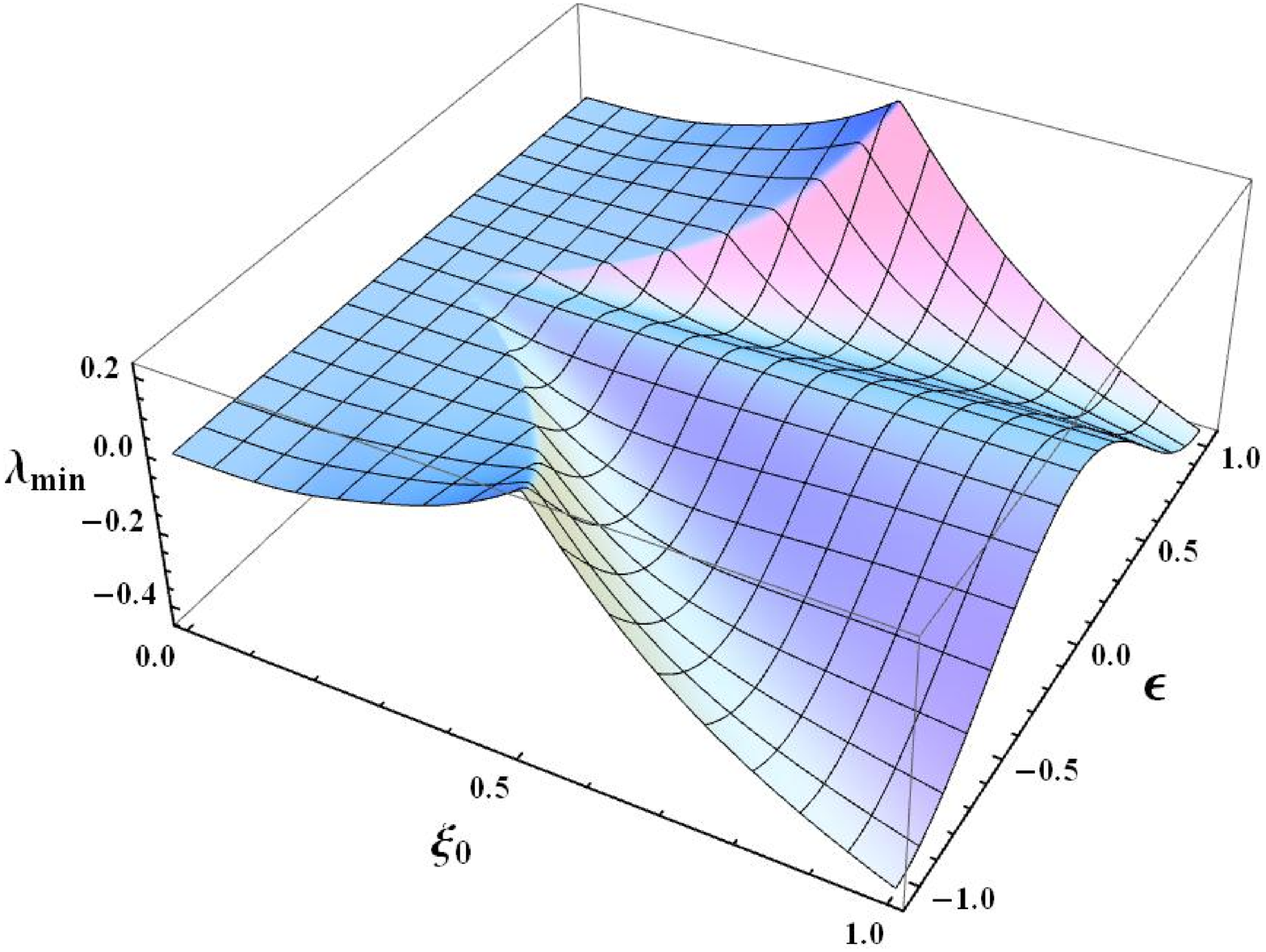}
\label{fig:minev3Dsmall}
}
\caption{Minimal eigenvalue $\lambda_{\rm min}$ of the partial transpose of the matrix of moments, $M^{\Gamma}$, for an entangled cat state $\ket{\psi_+}$ given in Eq.~\eqref{eq:EntangledCat} and for values of $\alpha_i$ and $\beta_i$ as defined in Eq.~\eqref{eq:alphabeta}.}
\end{figure}
Now let us consider the entangled state $\ket{\psi_+}$ of two nanomechanical resonators from Eq. (\ref{eq:EntangledCat}) in Sec. \ref{ss:prep}.
By carefully adjusting the interaction times $\tau_1$ and $\tau_2$ we can choose $\alpha_{1,2}$ and $\beta_{1,2}$ in order to realize the following settings
\bea
\alpha_1=2\xi_0,&\;\alpha_2=i\varepsilon/2\xi_0,\nonumber\\
\beta_1=-2\xi_0,&\;\beta_2=-i\varepsilon/2\xi_0,\label{eq:alphabeta}
\eea
with some real parameter $\varepsilon$. For these parameters we can now evaluate the matrix of moments~\eqref{eq:twooscmom} and calculate the minimal eigenvalue $\lambda_{\rm min}$ of the partially transposed matrix $M^{\Gamma}$. A negative value of $\lambda_{\rm min}$ certifies that the resonator state is entangled. As shown in Fig.~\ref{fig:minev3d} and Fig.~\ref{fig:minev3Dsmall} this occurs for the present example  for values of $\xi_0\gtrsim 0.3$ and $\varepsilon\gtrsim 0.1$, i.e., for quite modest values of the displacement amplitudes.

%

This entanglement criterion can also be used to construct a common entanglement witnesses. For instance, if $\vec \eta$ with elements $\eta_{ij}$ is an eigenvector that corresponds to a negative expectation value of the partially transposed matrix of moments, \textit{i.e.}, $\vec \eta^\dag M^{\Gamma} ~\vec \eta~<0$, then one has
\begin{equation}
\label{eq:entwit}
\tr{ \vec \eta \vec \eta^\dag M^{\Gamma}  } = \tr{ (\vec \eta \vec \eta^\dag)^{\Gamma} M}\equiv \tr{W\rho} < 0.
\end{equation}
In the last step of Eq.~\eqref{eq:entwit} we interpret the expression as a linear expectation value of an entanglement witness operator $W$ on the state $\rho$.
For a matrix of moments of the form $M_{ij,kl}=\tr{V_{ij}^\dag  V_{kl} \rho}$ the witness operator is then explicitly given as
\begin{equation}
W=\sum_{ij,kl} \eta_{kj} \eta_{il}^* V_{ij}^\dag   V_{kl}.
\end{equation}

\begin{figure}
\includegraphics[width=.95\columnwidth]{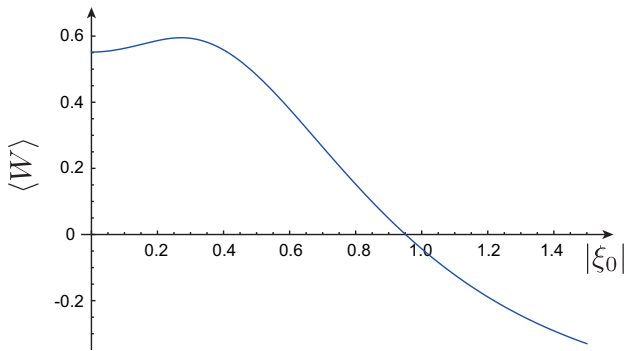}
\caption{Expectation value of the entanglement witness $W$ defined in Eq.~\eqref{eq:witness_example} for the state $\ket{\psi_+}$~\eqref{eq:EntangledCat}. The plot is shown for different values of the cat size $\xi_0$ and for fixed $\varepsilon=\pi/2$ and $w\approx0.4247$.}
\label{fig:witness}
\end{figure}

One can follow this approach for the example of the cat state in Eq. (\ref{eq:EntangledCat}). From empirical observations we find that for the ideal state~\eqref{eq:EntangledCat} and for $\xi_0\gtrsim 2 $, the eigenvector that corresponds to the minimal eigenvalue of the matrix $M^{\Gamma}$ has a particularly simple form
\begin{equation}
\vec \eta=[w, 0, -iw, 0, -\sqrt{1-4w^2},0,iw, 0, w]^T,
\end{equation}
with $w\approx0.43$ \footnote{Note that for smaller values of $\xi_0$ the vector $\vec \eta$ has another form and one can construct a witness that a little bit better than the presented one. However, it can not be described with only one parameter.}. The corresponding entanglement witness reads, using the abbreviations for three different measurement settings $s_1=2\xi_0,s_2=i\varepsilon/2\xi_0$ and $s_3=s_2-s_1$,
\begin{widetext}
\begin{align}
\label{eq:witness_example}
W =& \eins \otimes \eins + w^2\left\{[\mathcal{D}(s_2)-\mathcal{D}(-s_2)]\otimes[\mathcal{D}(s_2)-\mathcal{D}(-s_2)]+2i\left([\mathcal{D}(s_2)-\mathcal{D}(-s_2)]\otimes \eins -\eins\otimes [\mathcal{D}(s_2)-\mathcal{D}(-s_2)]\right)\right\} \\
\nonumber
&-w\sqrt{1-4w^2}\left\{ \mathcal{D}(s_1)\otimes \mathcal{D}(s_1)+\mathcal{D}(-s_1)\otimes \mathcal{D}(-s_1)+\mathcal{D}(s_3)\otimes \mathcal{D}(s_3)+\mathcal{D}(-s_3)\otimes \mathcal{D}(-s_3) \right. \\
\nonumber
& \:\:\:\:\:\:\:\:\:\:\:\:\:\:\:\:\:\:\:\:\:\:\:\:\:\:\:\:\:\:\:\left. i\ex{-i\varepsilon}[\mathcal{D}(s_1)\otimes\mathcal{D}(-s_3)+\mathcal{D}(s_3)\otimes\mathcal{D}(-s_1)]-i\ex{i\varepsilon}[\mathcal{D}(-s_1)\otimes\mathcal{D}(s_3)+\mathcal{D}(-s_3)\otimes\mathcal{D}(s_1)] \right\}.
\end{align}
\end{widetext}
We emphasize that this witness does not require the measurement  of all possible correlations. The first line can be measured alone by using $s_2$, while the remaining parts of the witness only include correlations between the $s_1$ and $s_3$ settings. Since most of the expectation values are complex conjugate to each other, the witness $W$ only requires the measurement  of eight independent correlations. As shown in Fig.~\ref{fig:witness} it still detects the entanglement of the $|\psi_+\rangle$ state for $|\xi_0|\gtrsim 1$.

\section{Conclusions}\label{sec:Conclusions}
In summary we have described a set of protocols for performing non-classicality tests and entanglement verification based on Ramsey-type measurement schemes for harmonic oscillators coupled to a two level system.  Specifically, by extending previous ideas for non-classicality test for single resonator modes, we have shown that the same underlying techniques can be used to verify the entanglement between two Schr\"odinger cat states. Although for the coupling under consideration this task cannot be achieved directly, we have identified a general strategy for constructing an entanglement witness for this problem, which then requires only a small set of correlation measurements and not the full knowledge of the two-mode Wigner function. This work is mainly motivated by upcoming experiments where nanomechanical resonators are strongly coupled to microscopic two-level systems, and where such techniques could be used to test the principles of superpositions and entanglement with massive objects.  However, the analysis of this work is quite general and can be applied to other systems, for example trapped ions, as well.

\section{Acknowledgements} This work was supported by the European Project SIQS, the WWTF and the
Austrian Science Fund (FWF) through SFB FOQUS and the START grant Y 591-N16, the BMBF (Chist-Era Project QUASAR),
the FQXi Fund (Silicon Valley Community Foundation) and the DFG.
$\rm OG^1$ and $\rm OG^2$ were also supported by the Marie Curie Actions (FWF Erwin Schr\"odinger Stipendium J3312-N27 and CIG 293993/ENFOQI respectively).
$\rm OG^1$ especially acknowledges the patience of his wife, because he has to finish this paper a long after he left the academia and worked on ot in the free time.

\begin{thebibliography}{61}
\expandafter\ifx\csname natexlab\endcsname\relax\def\natexlab#1{#1}\fi
\expandafter\ifx\csname bibnamefont\endcsname\relax
  \def\bibnamefont#1{#1}\fi
\expandafter\ifx\csname bibfnamefont\endcsname\relax
  \def\bibfnamefont#1{#1}\fi
\expandafter\ifx\csname citenamefont\endcsname\relax
  \def\citenamefont#1{#1}\fi
\expandafter\ifx\csname url\endcsname\relax
  \def\url#1{\texttt{#1}}\fi
\expandafter\ifx\csname urlprefix\endcsname\relax\def\urlprefix{URL }\fi
\providecommand{\bibinfo}[2]{#2}
\providecommand{\eprint}[2][]{\url{#2}}

\bibitem[{\citenamefont{Aspect}(1999)}]{Aspect1999}
\bibinfo{author}{\bibfnamefont{A.}~\bibnamefont{Aspect}},
  \bibinfo{journal}{Nature} \textbf{\bibinfo{volume}{398}},
  \bibinfo{pages}{189} (\bibinfo{year}{1999}).

\bibitem[{\citenamefont{Haroche}(2013)}]{Haroche2013}
\bibinfo{author}{\bibfnamefont{S.}~\bibnamefont{Haroche}},
  \bibinfo{journal}{Rev. Mod. Phys.} \textbf{\bibinfo{volume}{85}},
  \bibinfo{pages}{1083} (\bibinfo{year}{2013}).

\bibitem[{\citenamefont{Kirchmair et~al.}(2013)\citenamefont{Kirchmair,
  Vlastakis, Leghtas, Nigg, Paik, Ginossar, Mirrahimi, Frunzio, Girvin, and
  Schoelkopf}}]{Kirchmair2013}
\bibinfo{author}{\bibfnamefont{G.}~\bibnamefont{Kirchmair}},
  \bibinfo{author}{\bibfnamefont{B.}~\bibnamefont{Vlastakis}},
  \bibinfo{author}{\bibfnamefont{Z.}~\bibnamefont{Leghtas}},
  \bibinfo{author}{\bibfnamefont{S.~E.} \bibnamefont{Nigg}},
  \bibinfo{author}{\bibfnamefont{H.}~\bibnamefont{Paik}},
  \bibinfo{author}{\bibfnamefont{E.}~\bibnamefont{Ginossar}},
  \bibinfo{author}{\bibfnamefont{M.}~\bibnamefont{Mirrahimi}},
  \bibinfo{author}{\bibfnamefont{L.}~\bibnamefont{Frunzio}},
  \bibinfo{author}{\bibfnamefont{S.~M.} \bibnamefont{Girvin}},
  \bibnamefont{and} \bibinfo{author}{\bibfnamefont{R.~J.}
  \bibnamefont{Schoelkopf}}, \bibinfo{journal}{Nature}
  \textbf{\bibinfo{volume}{495}}, \bibinfo{pages}{205} (\bibinfo{year}{2013}).

\bibitem[{\citenamefont{Kasevich and Chu}(1991)}]{Kasevich1991}
\bibinfo{author}{\bibfnamefont{M.}~\bibnamefont{Kasevich}} \bibnamefont{and}
  \bibinfo{author}{\bibfnamefont{S.}~\bibnamefont{Chu}},
  \bibinfo{journal}{Phys. Rev. Lett.} \textbf{\bibinfo{volume}{67}},
  \bibinfo{pages}{181} (\bibinfo{year}{1991}).

\bibitem[{\citenamefont{Wineland}(2013)}]{WinelandRMP2013}
\bibinfo{author}{\bibfnamefont{D.~J.} \bibnamefont{Wineland}},
  \bibinfo{journal}{Rev. Mod. Phys.} \textbf{\bibinfo{volume}{85}},
  \bibinfo{pages}{1103} (\bibinfo{year}{2013}).

\bibitem[{\citenamefont{Monz et~al.}(2011)\citenamefont{Monz, Schindler,
  Barreiro, Chwalla, Nigg, Coish, Harlander, H\"ansel, Hennrich, and
  Blatt}}]{Monz2011}
\bibinfo{author}{\bibfnamefont{T.}~\bibnamefont{Monz}},
  \bibinfo{author}{\bibfnamefont{P.}~\bibnamefont{Schindler}},
  \bibinfo{author}{\bibfnamefont{J.~T.} \bibnamefont{Barreiro}},
  \bibinfo{author}{\bibfnamefont{M.}~\bibnamefont{Chwalla}},
  \bibinfo{author}{\bibfnamefont{D.}~\bibnamefont{Nigg}},
  \bibinfo{author}{\bibfnamefont{W.~A.} \bibnamefont{Coish}},
  \bibinfo{author}{\bibfnamefont{M.}~\bibnamefont{Harlander}},
  \bibinfo{author}{\bibfnamefont{W.}~\bibnamefont{H\"ansel}},
  \bibinfo{author}{\bibfnamefont{M.}~\bibnamefont{Hennrich}}, \bibnamefont{and}
  \bibinfo{author}{\bibfnamefont{R.}~\bibnamefont{Blatt}},
  \bibinfo{journal}{Phys. Rev. Lett.} \textbf{\bibinfo{volume}{106}},
  \bibinfo{pages}{130506} (\bibinfo{year}{2011}).

\bibitem[{\citenamefont{Robens et~al.}(2014)\citenamefont{Robens, Alt,
  Meschede, Emary, and Alberti}}]{Robens2014}
\bibinfo{author}{\bibfnamefont{C.}~\bibnamefont{Robens}},
  \bibinfo{author}{\bibfnamefont{W.}~\bibnamefont{Alt}},
  \bibinfo{author}{\bibfnamefont{D.}~\bibnamefont{Meschede}},
  \bibinfo{author}{\bibfnamefont{C.}~\bibnamefont{Emary}}, \bibnamefont{and}
  \bibinfo{author}{\bibfnamefont{A.}~\bibnamefont{Alberti}},
  \emph{\bibinfo{title}{Quantum diffusion falsifies the concept of classical
  trajectories by violation of leggett-garg inequality}},
  \bibinfo{howpublished}{arXiv:1404.3912} (\bibinfo{year}{2014}).

\bibitem[{\citenamefont{Arndt and Hornberger}(2014)}]{Arndt2014}
\bibinfo{author}{\bibfnamefont{M.}~\bibnamefont{Arndt}} \bibnamefont{and}
  \bibinfo{author}{\bibfnamefont{K.}~\bibnamefont{Hornberger}},
  \bibinfo{journal}{Nat Phys} \textbf{\bibinfo{volume}{10}},
  \bibinfo{pages}{271} (\bibinfo{year}{2014}).

\bibitem[{\citenamefont{Briggs et~al.}(2013)\citenamefont{Briggs, Butterfield,
  and Zeilinger}}]{OxfordQuestions}
\bibinfo{author}{\bibfnamefont{G.~A.~D.} \bibnamefont{Briggs}},
  \bibinfo{author}{\bibfnamefont{J.~N.} \bibnamefont{Butterfield}},
  \bibnamefont{and}
  \bibinfo{author}{\bibfnamefont{A.}~\bibnamefont{Zeilinger}},
  \bibinfo{journal}{Proceedings of the Royal Society A: Mathematical, Physical
  and Engineering Science} \textbf{\bibinfo{volume}{469}}
  (\bibinfo{year}{2013}).

\bibitem[{\citenamefont{Ghirardi et~al.}(1986)\citenamefont{Ghirardi, Rimini,
  and Weber}}]{GPW}
\bibinfo{author}{\bibfnamefont{G.~C.} \bibnamefont{Ghirardi}},
  \bibinfo{author}{\bibfnamefont{A.}~\bibnamefont{Rimini}}, \bibnamefont{and}
  \bibinfo{author}{\bibfnamefont{T.}~\bibnamefont{Weber}},
  \bibinfo{journal}{Phys. Rev. D} \textbf{\bibinfo{volume}{34}},
  \bibinfo{pages}{470} (\bibinfo{year}{1986}).

\bibitem[{\citenamefont{Di\'osi}(1989)}]{Diosi}
\bibinfo{author}{\bibfnamefont{L.}~\bibnamefont{Di\'osi}},
  \bibinfo{journal}{Phys. Rev. A} \textbf{\bibinfo{volume}{40}},
  \bibinfo{pages}{1165} (\bibinfo{year}{1989}).

\bibitem[{\citenamefont{Penrose}(1996)}]{Penrose}
\bibinfo{author}{\bibfnamefont{R.}~\bibnamefont{Penrose}},
  \bibinfo{journal}{General Relativity and Gravitation}
  \textbf{\bibinfo{volume}{28}}, \bibinfo{pages}{581} (\bibinfo{year}{1996}).

\bibitem[{\citenamefont{Ghirardi et~al.}(1990)\citenamefont{Ghirardi, Pearle,
  and Rimini}}]{GhirardiPRA1990}
\bibinfo{author}{\bibfnamefont{G.~C.} \bibnamefont{Ghirardi}},
  \bibinfo{author}{\bibfnamefont{P.}~\bibnamefont{Pearle}}, \bibnamefont{and}
  \bibinfo{author}{\bibfnamefont{A.}~\bibnamefont{Rimini}},
  \bibinfo{journal}{Phys. Rev. A} \textbf{\bibinfo{volume}{42}},
  \bibinfo{pages}{78} (\bibinfo{year}{1990}).

\bibitem[{\citenamefont{Bassi et~al.}(2013)\citenamefont{Bassi, Lochan, Satin,
  Singh, and Ulbricht}}]{Bassi2013}
\bibinfo{author}{\bibfnamefont{A.}~\bibnamefont{Bassi}},
  \bibinfo{author}{\bibfnamefont{K.}~\bibnamefont{Lochan}},
  \bibinfo{author}{\bibfnamefont{S.}~\bibnamefont{Satin}},
  \bibinfo{author}{\bibfnamefont{T.~P.} \bibnamefont{Singh}}, \bibnamefont{and}
  \bibinfo{author}{\bibfnamefont{H.}~\bibnamefont{Ulbricht}},
  \bibinfo{journal}{Rev. Mod. Phys.} \textbf{\bibinfo{volume}{85}},
  \bibinfo{pages}{471} (\bibinfo{year}{2013}).

\bibitem[{\citenamefont{O'Connell et~al.}(2010)\citenamefont{O'Connell,
  Hofheinz, Ansmann, Bialczak, Lenander, Lucero, Neeley, Sank, Wang, Weides
  et~al.}}]{OConnel2010}
\bibinfo{author}{\bibfnamefont{A.~D.} \bibnamefont{O'Connell}},
  \bibinfo{author}{\bibfnamefont{M.}~\bibnamefont{Hofheinz}},
  \bibinfo{author}{\bibfnamefont{M.}~\bibnamefont{Ansmann}},
  \bibinfo{author}{\bibfnamefont{R.~C.} \bibnamefont{Bialczak}},
  \bibinfo{author}{\bibfnamefont{M.}~\bibnamefont{Lenander}},
  \bibinfo{author}{\bibfnamefont{E.}~\bibnamefont{Lucero}},
  \bibinfo{author}{\bibfnamefont{M.}~\bibnamefont{Neeley}},
  \bibinfo{author}{\bibfnamefont{D.}~\bibnamefont{Sank}},
  \bibinfo{author}{\bibfnamefont{H.}~\bibnamefont{Wang}},
  \bibinfo{author}{\bibfnamefont{M.}~\bibnamefont{Weides}},
  \bibnamefont{et~al.}, \bibinfo{journal}{Nature}
  \textbf{\bibinfo{volume}{464}}, \bibinfo{pages}{697} (\bibinfo{year}{2010}).

\bibitem[{\citenamefont{Teufel et~al.}(2011)\citenamefont{Teufel, Donner, Li,
  Harlow, Allman, Cicak, Sirois, Whittaker, Lehnert, and
  Simmonds}}]{Teufel2011}
\bibinfo{author}{\bibfnamefont{J.~D.} \bibnamefont{Teufel}},
  \bibinfo{author}{\bibfnamefont{T.}~\bibnamefont{Donner}},
  \bibinfo{author}{\bibfnamefont{D.}~\bibnamefont{Li}},
  \bibinfo{author}{\bibfnamefont{J.~W.} \bibnamefont{Harlow}},
  \bibinfo{author}{\bibfnamefont{M.~S.} \bibnamefont{Allman}},
  \bibinfo{author}{\bibfnamefont{K.}~\bibnamefont{Cicak}},
  \bibinfo{author}{\bibfnamefont{A.~J.} \bibnamefont{Sirois}},
  \bibinfo{author}{\bibfnamefont{J.~D.} \bibnamefont{Whittaker}},
  \bibinfo{author}{\bibfnamefont{K.~W.} \bibnamefont{Lehnert}},
  \bibnamefont{and} \bibinfo{author}{\bibfnamefont{R.~W.}
  \bibnamefont{Simmonds}}, \bibinfo{journal}{Nature}
  \textbf{\bibinfo{volume}{475}}, \bibinfo{pages}{359} (\bibinfo{year}{2011}).

\bibitem[{\citenamefont{Chan et~al.}(2011)\citenamefont{Chan, Alegre,
  Safavi-Naeini, Hill, Krause, Groblacher, Aspelmeyer, and Painter}}]{Chan2011}
\bibinfo{author}{\bibfnamefont{J.}~\bibnamefont{Chan}},
  \bibinfo{author}{\bibfnamefont{T.~P.~M.} \bibnamefont{Alegre}},
  \bibinfo{author}{\bibfnamefont{A.~H.} \bibnamefont{Safavi-Naeini}},
  \bibinfo{author}{\bibfnamefont{J.~T.} \bibnamefont{Hill}},
  \bibinfo{author}{\bibfnamefont{A.}~\bibnamefont{Krause}},
  \bibinfo{author}{\bibfnamefont{S.}~\bibnamefont{Groblacher}},
  \bibinfo{author}{\bibfnamefont{M.}~\bibnamefont{Aspelmeyer}},
  \bibnamefont{and} \bibinfo{author}{\bibfnamefont{O.}~\bibnamefont{Painter}},
  \bibinfo{journal}{Nature} \textbf{\bibinfo{volume}{478}}, \bibinfo{pages}{89}
  (\bibinfo{year}{2011}).

\bibitem[{\citenamefont{Palomaki et~al.}(2013)\citenamefont{Palomaki, Teufel,
  Simmonds, and Lehnert}}]{Palomaki2013}
\bibinfo{author}{\bibfnamefont{T.~A.} \bibnamefont{Palomaki}},
  \bibinfo{author}{\bibfnamefont{J.~D.} \bibnamefont{Teufel}},
  \bibinfo{author}{\bibfnamefont{R.~W.} \bibnamefont{Simmonds}},
  \bibnamefont{and} \bibinfo{author}{\bibfnamefont{K.~W.}
  \bibnamefont{Lehnert}}, \bibinfo{journal}{Science}
  \textbf{\bibinfo{volume}{342}}, \bibinfo{pages}{710} (\bibinfo{year}{2013}).

\bibitem[{\citenamefont{Arcizet et~al.}(2011)\citenamefont{Arcizet, Jacques,
  Siria, Poncharal, Vincent, and Seidelin}}]{ArcizetNatPhys2011}
\bibinfo{author}{\bibfnamefont{O.}~\bibnamefont{Arcizet}},
  \bibinfo{author}{\bibfnamefont{V.}~\bibnamefont{Jacques}},
  \bibinfo{author}{\bibfnamefont{A.}~\bibnamefont{Siria}},
  \bibinfo{author}{\bibfnamefont{P.}~\bibnamefont{Poncharal}},
  \bibinfo{author}{\bibfnamefont{P.}~\bibnamefont{Vincent}}, \bibnamefont{and}
  \bibinfo{author}{\bibfnamefont{S.}~\bibnamefont{Seidelin}},
  \bibinfo{journal}{Nat Phys} \textbf{\bibinfo{volume}{7}},
  \bibinfo{pages}{879} (\bibinfo{year}{2011}).

\bibitem[{\citenamefont{Kolkowitz et~al.}(2012)\citenamefont{Kolkowitz,
  Bleszynski~Jayich, Unterreithmeier, Bennett, Rabl, Harris, and
  Lukin}}]{KolkowitzScience2012}
\bibinfo{author}{\bibfnamefont{S.}~\bibnamefont{Kolkowitz}},
  \bibinfo{author}{\bibfnamefont{A.~C.} \bibnamefont{Bleszynski~Jayich}},
  \bibinfo{author}{\bibfnamefont{Q.~P.} \bibnamefont{Unterreithmeier}},
  \bibinfo{author}{\bibfnamefont{S.~D.} \bibnamefont{Bennett}},
  \bibinfo{author}{\bibfnamefont{P.}~\bibnamefont{Rabl}},
  \bibinfo{author}{\bibfnamefont{J.~G.~E.} \bibnamefont{Harris}},
  \bibnamefont{and} \bibinfo{author}{\bibfnamefont{M.~D.} \bibnamefont{Lukin}},
  \bibinfo{journal}{Science} \textbf{\bibinfo{volume}{335}},
  \bibinfo{pages}{1603} (\bibinfo{year}{2012}).

\bibitem[{\citenamefont{Ovartchaiyapong
  et~al.}(2014)\citenamefont{Ovartchaiyapong, Lee, Myers, and
  Jayich}}]{Ovartchaiyapong2014}
\bibinfo{author}{\bibfnamefont{P.}~\bibnamefont{Ovartchaiyapong}},
  \bibinfo{author}{\bibfnamefont{K.~W.} \bibnamefont{Lee}},
  \bibinfo{author}{\bibfnamefont{B.~A.} \bibnamefont{Myers}}, \bibnamefont{and}
  \bibinfo{author}{\bibfnamefont{A.~C.~B.} \bibnamefont{Jayich}},
  \bibinfo{journal}{Nat. Comm.} \textbf{\bibinfo{volume}{5}}
  (\bibinfo{year}{2014}).

\bibitem[{\citenamefont{Teissier et~al.}(2014)\citenamefont{Teissier, Barfuss,
  Appel, Neu, and Maletinsky}}]{TeissierPRL2014}
\bibinfo{author}{\bibfnamefont{J.}~\bibnamefont{Teissier}},
  \bibinfo{author}{\bibfnamefont{A.}~\bibnamefont{Barfuss}},
  \bibinfo{author}{\bibfnamefont{P.}~\bibnamefont{Appel}},
  \bibinfo{author}{\bibfnamefont{E.}~\bibnamefont{Neu}}, \bibnamefont{and}
  \bibinfo{author}{\bibfnamefont{P.}~\bibnamefont{Maletinsky}},
  \bibinfo{journal}{Phys. Rev. Lett.} \textbf{\bibinfo{volume}{113}},
  \bibinfo{pages}{020503} (\bibinfo{year}{2014}).

\bibitem[{\citenamefont{LaHaye et~al.}(2009)\citenamefont{LaHaye, Suh,
  Echternach, Schwab, and Roukes}}]{LaHayeNature2009}
\bibinfo{author}{\bibfnamefont{M.~D.} \bibnamefont{LaHaye}},
  \bibinfo{author}{\bibfnamefont{J.}~\bibnamefont{Suh}},
  \bibinfo{author}{\bibfnamefont{P.~M.} \bibnamefont{Echternach}},
  \bibinfo{author}{\bibfnamefont{K.~C.} \bibnamefont{Schwab}},
  \bibnamefont{and} \bibinfo{author}{\bibfnamefont{M.~L.}
  \bibnamefont{Roukes}}, \bibinfo{journal}{Nature}
  \textbf{\bibinfo{volume}{459}}, \bibinfo{pages}{960} (\bibinfo{year}{2009}).

\bibitem[{\citenamefont{Pirkkalainen et~al.}(2013)\citenamefont{Pirkkalainen,
  Cho, Li, Paraoanu, Hakonen, and Sillanpaa}}]{PirkkalainenNature2013}
\bibinfo{author}{\bibfnamefont{J.-M.} \bibnamefont{Pirkkalainen}},
  \bibinfo{author}{\bibfnamefont{S.~U.} \bibnamefont{Cho}},
  \bibinfo{author}{\bibfnamefont{J.}~\bibnamefont{Li}},
  \bibinfo{author}{\bibfnamefont{G.~S.} \bibnamefont{Paraoanu}},
  \bibinfo{author}{\bibfnamefont{P.~J.} \bibnamefont{Hakonen}},
  \bibnamefont{and} \bibinfo{author}{\bibfnamefont{M.~A.}
  \bibnamefont{Sillanpaa}}, \bibinfo{journal}{Nature}
  \textbf{\bibinfo{volume}{494}}, \bibinfo{pages}{211} (\bibinfo{year}{2013}).

\bibitem[{\citenamefont{Poot and van~der Zant}(2012)}]{Poot2012}
\bibinfo{author}{\bibfnamefont{M.}~\bibnamefont{Poot}} \bibnamefont{and}
  \bibinfo{author}{\bibfnamefont{H.~S.} \bibnamefont{van~der Zant}},
  \bibinfo{journal}{Phys. Rep.} \textbf{\bibinfo{volume}{511}},
  \bibinfo{pages}{273 } (\bibinfo{year}{2012}).

\bibitem[{\citenamefont{Aspelmeyer et~al.}(2013)\citenamefont{Aspelmeyer,
  Kippenberg, and Marquardt}}]{Aspelmeyer2013}
\bibinfo{author}{\bibfnamefont{M.}~\bibnamefont{Aspelmeyer}},
  \bibinfo{author}{\bibfnamefont{T.~J.} \bibnamefont{Kippenberg}},
  \bibnamefont{and}
  \bibinfo{author}{\bibfnamefont{F.}~\bibnamefont{Marquardt}},
  \emph{\bibinfo{title}{Cavity optomechanics}},
  \bibinfo{howpublished}{arXiv:1303.0733} (\bibinfo{year}{2013}).

\bibitem[{\citenamefont{Armour et~al.}(2002)\citenamefont{Armour, Blencowe, and
  Schwab}}]{ArmourPRL2002}
\bibinfo{author}{\bibfnamefont{A.~D.} \bibnamefont{Armour}},
  \bibinfo{author}{\bibfnamefont{M.~P.} \bibnamefont{Blencowe}},
  \bibnamefont{and} \bibinfo{author}{\bibfnamefont{K.~C.}
  \bibnamefont{Schwab}}, \bibinfo{journal}{Phys. Rev. Lett.}
  \textbf{\bibinfo{volume}{88}}, \bibinfo{pages}{148301}
  (\bibinfo{year}{2002}).

\bibitem[{\citenamefont{Tian}(2005)}]{TianPRB2005}
\bibinfo{author}{\bibfnamefont{L.}~\bibnamefont{Tian}}, \bibinfo{journal}{Phys.
  Rev. B} \textbf{\bibinfo{volume}{72}}, \bibinfo{pages}{195411}
  (\bibinfo{year}{2005}).

\bibitem[{\citenamefont{Marshall et~al.}(2003)\citenamefont{Marshall, Simon,
  Penrose, and Bouwmeester}}]{MarshallPRL2003}
\bibinfo{author}{\bibfnamefont{W.}~\bibnamefont{Marshall}},
  \bibinfo{author}{\bibfnamefont{C.}~\bibnamefont{Simon}},
  \bibinfo{author}{\bibfnamefont{R.}~\bibnamefont{Penrose}}, \bibnamefont{and}
  \bibinfo{author}{\bibfnamefont{D.}~\bibnamefont{Bouwmeester}},
  \bibinfo{journal}{Phys. Rev. Lett.} \textbf{\bibinfo{volume}{91}},
  \bibinfo{pages}{130401} (\bibinfo{year}{2003}).

\bibitem[{\citenamefont{Asadian et~al.}(2014)\citenamefont{Asadian, Brukner,
  and Rabl}}]{asadian_probing_2013}
\bibinfo{author}{\bibfnamefont{A.}~\bibnamefont{Asadian}},
  \bibinfo{author}{\bibfnamefont{C.}~\bibnamefont{Brukner}}, \bibnamefont{and}
  \bibinfo{author}{\bibfnamefont{P.}~\bibnamefont{Rabl}},
  \bibinfo{journal}{Phys. Rev. Lett.} \textbf{\bibinfo{volume}{112}},
  \bibinfo{pages}{190402} (\bibinfo{year}{2014}).

\bibitem[{\citenamefont{Leggett and Garg}(1985)}]{LG85}
\bibinfo{author}{\bibfnamefont{A.~J.} \bibnamefont{Leggett}} \bibnamefont{and}
  \bibinfo{author}{\bibfnamefont{A.}~\bibnamefont{Garg}},
  \bibinfo{journal}{Phys. Rev. Lett.} \textbf{\bibinfo{volume}{54}},
  \bibinfo{pages}{857} (\bibinfo{year}{1985}).

\bibitem[{\citenamefont{Emary et~al.}(2014)\citenamefont{Emary, Lambert, and
  Nori}}]{Nori}
\bibinfo{author}{\bibfnamefont{C.}~\bibnamefont{Emary}},
  \bibinfo{author}{\bibfnamefont{N.}~\bibnamefont{Lambert}}, \bibnamefont{and}
  \bibinfo{author}{\bibfnamefont{F.}~\bibnamefont{Nori}},
  \bibinfo{journal}{Reports on Progress in Physics}
  \textbf{\bibinfo{volume}{77}}, \bibinfo{pages}{016001}
  (\bibinfo{year}{2014}).

\bibitem[{\citenamefont{Rabl et~al.}(2004)\citenamefont{Rabl, Shnirman, and
  Zoller}}]{Rabl2004}
\bibinfo{author}{\bibfnamefont{P.}~\bibnamefont{Rabl}},
  \bibinfo{author}{\bibfnamefont{A.}~\bibnamefont{Shnirman}}, \bibnamefont{and}
  \bibinfo{author}{\bibfnamefont{P.}~\bibnamefont{Zoller}},
  \bibinfo{journal}{Phys. Rev. B} \textbf{\bibinfo{volume}{70}},
  \bibinfo{pages}{205304} (\bibinfo{year}{2004}).

\bibitem[{\citenamefont{Zhang et~al.}(2006)\citenamefont{Zhang, Xie, and
  Yang}}]{Zhang2006}
\bibinfo{author}{\bibfnamefont{S.-Z.} \bibnamefont{Zhang}},
  \bibinfo{author}{\bibfnamefont{X.-T.} \bibnamefont{Xie}}, \bibnamefont{and}
  \bibinfo{author}{\bibfnamefont{W.-X.} \bibnamefont{Yang}},
  \bibinfo{journal}{Communications in Theoretical Physics}
  \textbf{\bibinfo{volume}{46}}, \bibinfo{pages}{306} (\bibinfo{year}{2006}).

\bibitem[{\citenamefont{Singh and Meystre}(2010)}]{Singh2010}
\bibinfo{author}{\bibfnamefont{S.}~\bibnamefont{Singh}} \bibnamefont{and}
  \bibinfo{author}{\bibfnamefont{P.}~\bibnamefont{Meystre}},
  \bibinfo{journal}{Phys. Rev. A} \textbf{\bibinfo{volume}{81}},
  \bibinfo{pages}{041804} (\bibinfo{year}{2010}).

\bibitem[{\citenamefont{Tufarelli et~al.}(2011)\citenamefont{Tufarelli, Kim,
  and Bose}}]{Tufarelli2011}
\bibinfo{author}{\bibfnamefont{T.}~\bibnamefont{Tufarelli}},
  \bibinfo{author}{\bibfnamefont{M.~S.} \bibnamefont{Kim}}, \bibnamefont{and}
  \bibinfo{author}{\bibfnamefont{S.}~\bibnamefont{Bose}},
  \bibinfo{journal}{Phys. Rev. A} \textbf{\bibinfo{volume}{83}},
  \bibinfo{pages}{062120} (\bibinfo{year}{2011}).

\bibitem[{\citenamefont{Vogel}(2000)}]{vogel_nonclassical_2000}
\bibinfo{author}{\bibfnamefont{W.}~\bibnamefont{Vogel}},
  \bibinfo{journal}{Phys. Rev. Lett.} \textbf{\bibinfo{volume}{84}},
  \bibinfo{pages}{1849} (\bibinfo{year}{2000}).

\bibitem[{\citenamefont{Agarwal and Eberly}(2012)}]{Agarwal2012}
\bibinfo{author}{\bibfnamefont{S.}~\bibnamefont{Agarwal}} \bibnamefont{and}
  \bibinfo{author}{\bibfnamefont{J.~H.} \bibnamefont{Eberly}},
  \bibinfo{journal}{Phys. Rev. A} \textbf{\bibinfo{volume}{86}},
  \bibinfo{pages}{022341} (\bibinfo{year}{2012}).

\bibitem[{\citenamefont{Rabl et~al.}(2009)\citenamefont{Rabl, Cappellaro,
  Gurudev~Dutt, Jiang, Maze, and Lukin}}]{Rabl2009}
\bibinfo{author}{\bibfnamefont{P.}~\bibnamefont{Rabl}},
  \bibinfo{author}{\bibfnamefont{P.}~\bibnamefont{Cappellaro}},
  \bibinfo{author}{\bibfnamefont{M.~V.} \bibnamefont{Gurudev~Dutt}},
  \bibinfo{author}{\bibfnamefont{L.}~\bibnamefont{Jiang}},
  \bibinfo{author}{\bibfnamefont{J.~R.} \bibnamefont{Maze}}, \bibnamefont{and}
  \bibinfo{author}{\bibfnamefont{M.~D.} \bibnamefont{Lukin}},
  \bibinfo{journal}{Phys. Rev. B} \textbf{\bibinfo{volume}{79}},
  \bibinfo{pages}{041302} (\bibinfo{year}{2009}).

\bibitem[{\citenamefont{Wilson-Rae et~al.}(2004)\citenamefont{Wilson-Rae,
  Zoller, and Imamoglu}}]{WilsonRaePRL2004}
\bibinfo{author}{\bibfnamefont{I.}~\bibnamefont{Wilson-Rae}},
  \bibinfo{author}{\bibfnamefont{P.}~\bibnamefont{Zoller}}, \bibnamefont{and}
  \bibinfo{author}{\bibfnamefont{A.}~\bibnamefont{Imamoglu}},
  \bibinfo{journal}{Phys. Rev. Lett.} \textbf{\bibinfo{volume}{9}},
  \bibinfo{pages}{075507} (\bibinfo{year}{2004}).

\bibitem[{\citenamefont{Yeo et~al.}(2014)\citenamefont{Yeo, de~Assis, Gloppe,
  Dupont-Ferrier, Verlot, Malik, Dupuy, Claudon, Gerard, Auffeves
  et~al.}}]{Yeo2014}
\bibinfo{author}{\bibfnamefont{I.}~\bibnamefont{Yeo}},
  \bibinfo{author}{\bibfnamefont{P.-L.} \bibnamefont{de~Assis}},
  \bibinfo{author}{\bibfnamefont{A.}~\bibnamefont{Gloppe}},
  \bibinfo{author}{\bibfnamefont{E.}~\bibnamefont{Dupont-Ferrier}},
  \bibinfo{author}{\bibfnamefont{P.}~\bibnamefont{Verlot}},
  \bibinfo{author}{\bibfnamefont{N.~S.} \bibnamefont{Malik}},
  \bibinfo{author}{\bibfnamefont{E.}~\bibnamefont{Dupuy}},
  \bibinfo{author}{\bibfnamefont{J.}~\bibnamefont{Claudon}},
  \bibinfo{author}{\bibfnamefont{J.-M.} \bibnamefont{Gerard}},
  \bibinfo{author}{\bibfnamefont{A.}~\bibnamefont{Auffeves}},
  \bibnamefont{et~al.}, \bibinfo{journal}{Nature Nano.}
  \textbf{\bibinfo{volume}{9}}, \bibinfo{pages}{106} (\bibinfo{year}{2014}).

\bibitem[{\citenamefont{Montinaro et~al.}(2014)\citenamefont{Montinaro, W\"ust,
  Fontana, Russo-Averchi, Heiss, Fontcuberta~i Morral, Warburton, and
  Poggio}}]{Montinaro2014}
\bibinfo{author}{\bibfnamefont{M.}~\bibnamefont{Montinaro}},
  \bibinfo{author}{\bibfnamefont{G.~M.} \bibnamefont{W\"ust},
  \bibfnamefont{Munsch}},
  \bibinfo{author}{\bibfnamefont{Y.}~\bibnamefont{Fontana}},
  \bibinfo{author}{\bibfnamefont{E.}~\bibnamefont{Russo-Averchi}},
  \bibinfo{author}{\bibfnamefont{M.}~\bibnamefont{Heiss}},
  \bibinfo{author}{\bibfnamefont{A.}~\bibnamefont{Fontcuberta~i Morral}},
  \bibinfo{author}{\bibfnamefont{R.~J.} \bibnamefont{Warburton}},
  \bibnamefont{and} \bibinfo{author}{\bibfnamefont{M.}~\bibnamefont{Poggio}},
  \bibinfo{journal}{Nano Lett.} \textbf{\bibinfo{volume}{14}},
  \bibinfo{pages}{4454} (\bibinfo{year}{2014}).

\bibitem[{\citenamefont{Steinke et~al.}(2011)\citenamefont{Steinke, Singh,
  Tasgin, Meystre, Schwab, and Vengalattore}}]{SteinkePRA2011}
\bibinfo{author}{\bibfnamefont{S.~K.} \bibnamefont{Steinke}},
  \bibinfo{author}{\bibfnamefont{S.}~\bibnamefont{Singh}},
  \bibinfo{author}{\bibfnamefont{M.~E.} \bibnamefont{Tasgin}},
  \bibinfo{author}{\bibfnamefont{P.}~\bibnamefont{Meystre}},
  \bibinfo{author}{\bibfnamefont{K.~C.} \bibnamefont{Schwab}},
  \bibnamefont{and}
  \bibinfo{author}{\bibfnamefont{M.}~\bibnamefont{Vengalattore}},
  \bibinfo{journal}{Phys. Rev. A} \textbf{\bibinfo{volume}{84}},
  \bibinfo{pages}{023841} (\bibinfo{year}{2011}).

\bibitem[{\citenamefont{Bennett et~al.}(2012)\citenamefont{Bennett, Kolkowitz,
  Unterreithmeier, Rabl, Jayich, Harris, and Lukin}}]{BennettNJP2012}
\bibinfo{author}{\bibfnamefont{S.~D.} \bibnamefont{Bennett}},
  \bibinfo{author}{\bibfnamefont{S.}~\bibnamefont{Kolkowitz}},
  \bibinfo{author}{\bibfnamefont{Q.~P.} \bibnamefont{Unterreithmeier}},
  \bibinfo{author}{\bibfnamefont{P.}~\bibnamefont{Rabl}},
  \bibinfo{author}{\bibfnamefont{A.~C.~B.} \bibnamefont{Jayich}},
  \bibinfo{author}{\bibfnamefont{J.~G.~E.} \bibnamefont{Harris}},
  \bibnamefont{and} \bibinfo{author}{\bibfnamefont{M.~D.} \bibnamefont{Lukin}},
  \bibinfo{journal}{New Journal of Physics} \textbf{\bibinfo{volume}{14}},
  \bibinfo{pages}{125004} (\bibinfo{year}{2012}).

\bibitem[{\citenamefont{Qu et~al.}(2014)\citenamefont{Qu, Dong, Wang, and
  Agarwal}}]{Qu2014}
\bibinfo{author}{\bibfnamefont{K.}~\bibnamefont{Qu}},
  \bibinfo{author}{\bibfnamefont{C.}~\bibnamefont{Dong}},
  \bibinfo{author}{\bibfnamefont{H.}~\bibnamefont{Wang}}, \bibnamefont{and}
  \bibinfo{author}{\bibfnamefont{G.~S.} \bibnamefont{Agarwal}},
  \emph{\bibinfo{title}{Optomechanical ramsey interferometry}},
  \bibinfo{howpublished}{arXiv:1408.5305} (\bibinfo{year}{2014}).

\bibitem[{\citenamefont{Vacanti et~al.}(2013)\citenamefont{Vacanti,
  Paternostro, Palma, Kim, and Vedral}}]{VacantiPRA2013}
\bibinfo{author}{\bibfnamefont{G.}~\bibnamefont{Vacanti}},
  \bibinfo{author}{\bibfnamefont{M.}~\bibnamefont{Paternostro}},
  \bibinfo{author}{\bibfnamefont{G.~M.} \bibnamefont{Palma}},
  \bibinfo{author}{\bibfnamefont{M.~S.} \bibnamefont{Kim}}, \bibnamefont{and}
  \bibinfo{author}{\bibfnamefont{V.}~\bibnamefont{Vedral}},
  \bibinfo{journal}{Phys. Rev. A} \textbf{\bibinfo{volume}{88}},
  \bibinfo{pages}{013851} (\bibinfo{year}{2013}).

\bibitem[{\citenamefont{Aharonov and Rohrlich}(2005)}]{aharonovrohrlichBook}
\bibinfo{author}{\bibfnamefont{Y.}~\bibnamefont{Aharonov}} \bibnamefont{and}
  \bibinfo{author}{\bibfnamefont{D.}~\bibnamefont{Rohrlich}},
  \emph{\bibinfo{title}{Quantum Paradoxes}} (\bibinfo{publisher}{Wiley-VCH},
  \bibinfo{year}{2005}).

\bibitem[{\citenamefont{Titulaer and Glauber}(1965)}]{Titulaer1965}
\bibinfo{author}{\bibfnamefont{U.~M.} \bibnamefont{Titulaer}} \bibnamefont{and}
  \bibinfo{author}{\bibfnamefont{R.~J.} \bibnamefont{Glauber}},
  \bibinfo{journal}{Phys. Rev.} \textbf{\bibinfo{volume}{140}},
  \bibinfo{pages}{B676} (\bibinfo{year}{1965}).

\bibitem[{\citenamefont{Mandel}(1986)}]{Mandel1986}
\bibinfo{author}{\bibfnamefont{L.}~\bibnamefont{Mandel}},
  \bibinfo{journal}{Phys. Scr.} \textbf{\bibinfo{volume}{T12}},
  \bibinfo{pages}{34} (\bibinfo{year}{1986}).

\bibitem[{\citenamefont{Di\'osi}(2000)}]{Diosi2000}
\bibinfo{author}{\bibfnamefont{L.}~\bibnamefont{Di\'osi}},
  \bibinfo{journal}{Phys. Rev. Lett.} \textbf{\bibinfo{volume}{85}},
  \bibinfo{pages}{2841} (\bibinfo{year}{2000}).

\bibitem[{\citenamefont{Richter and
  Vogel}(2002)}]{richter_nonclassicality_2002}
\bibinfo{author}{\bibfnamefont{T.}~\bibnamefont{Richter}} \bibnamefont{and}
  \bibinfo{author}{\bibfnamefont{W.}~\bibnamefont{Vogel}},
  \bibinfo{journal}{Phys. Rev. Lett.} \textbf{\bibinfo{volume}{89}},
  \bibinfo{pages}{283601} (\bibinfo{year}{2002}).

\bibitem[{\citenamefont{Holevo}(2011)}]{HolevoBook11}
\bibinfo{author}{\bibfnamefont{A.}~\bibnamefont{Holevo}},
  \emph{\bibinfo{title}{Probabilistic and Statistical Aspects of Quantum
  Theory}} (\bibinfo{publisher}{Edizioni della Normale}, \bibinfo{year}{2011}).

\bibitem[{\citenamefont{Paavola et~al.}(2011)\citenamefont{Paavola, Hall,
  Paris, and Maniscalco}}]{Paavola2011}
\bibinfo{author}{\bibfnamefont{J.}~\bibnamefont{Paavola}},
  \bibinfo{author}{\bibfnamefont{M.~J.~W.} \bibnamefont{Hall}},
  \bibinfo{author}{\bibfnamefont{M.~G.~A.} \bibnamefont{Paris}},
  \bibnamefont{and}
  \bibinfo{author}{\bibfnamefont{S.}~\bibnamefont{Maniscalco}},
  \bibinfo{journal}{Phys. Rev. A} \textbf{\bibinfo{volume}{84}},
  \bibinfo{pages}{012121} (\bibinfo{year}{2011}).

\bibitem[{\citenamefont{Davidovich et~al.}(1993)\citenamefont{Davidovich,
  Maali, Brune, Raimond, and Haroche}}]{Davidovich1993}
\bibinfo{author}{\bibfnamefont{L.}~\bibnamefont{Davidovich}},
  \bibinfo{author}{\bibfnamefont{A.}~\bibnamefont{Maali}},
  \bibinfo{author}{\bibfnamefont{M.}~\bibnamefont{Brune}},
  \bibinfo{author}{\bibfnamefont{J.~M.} \bibnamefont{Raimond}},
  \bibnamefont{and} \bibinfo{author}{\bibfnamefont{S.}~\bibnamefont{Haroche}},
  \bibinfo{journal}{Phys. Rev. Lett.} \textbf{\bibinfo{volume}{71}},
  \bibinfo{pages}{2360} (\bibinfo{year}{1993}).

\bibitem[{\citenamefont{Hartmann et~al.}(2005)\citenamefont{Hartmann,
  Calsamiglia, D\"ur, and Briegel}}]{Walder-Hartmann_2005}
\bibinfo{author}{\bibfnamefont{L.}~\bibnamefont{Hartmann}},
  \bibinfo{author}{\bibfnamefont{J.}~\bibnamefont{Calsamiglia}},
  \bibinfo{author}{\bibfnamefont{W.}~\bibnamefont{D\"ur}}, \bibnamefont{and}
  \bibinfo{author}{\bibfnamefont{H.-J.} \bibnamefont{Briegel}},
  \bibinfo{journal}{Phys. Rev. A} \textbf{\bibinfo{volume}{72}},
  \bibinfo{pages}{052107} (\bibinfo{year}{2005}).

\bibitem[{\citenamefont{Shchukin and
  Vogel}(2005)}]{shchukin_inseparability_2005}
\bibinfo{author}{\bibfnamefont{E.}~\bibnamefont{Shchukin}} \bibnamefont{and}
  \bibinfo{author}{\bibfnamefont{W.}~\bibnamefont{Vogel}},
  \bibinfo{journal}{Phys. Rev. Lett.} \textbf{\bibinfo{volume}{95}},
  \bibinfo{pages}{230502} (\bibinfo{year}{2005}).

\bibitem[{\citenamefont{Miranowicz et~al.}(2009)\citenamefont{Miranowicz,
  Piani, Horodecki, and Horodecki}}]{miranowicz_inseparability_2009}
\bibinfo{author}{\bibfnamefont{A.}~\bibnamefont{Miranowicz}},
  \bibinfo{author}{\bibfnamefont{M.}~\bibnamefont{Piani}},
  \bibinfo{author}{\bibfnamefont{P.}~\bibnamefont{Horodecki}},
  \bibnamefont{and}
  \bibinfo{author}{\bibfnamefont{R.}~\bibnamefont{Horodecki}},
  \bibinfo{journal}{Phys. Rev. A} \textbf{\bibinfo{volume}{80}},
  \bibinfo{pages}{052303} (\bibinfo{year}{2009}).

\bibitem[{\citenamefont{Richter and Vogel}(2007)}]{richter_nonclassical_2007}
\bibinfo{author}{\bibfnamefont{T.}~\bibnamefont{Richter}} \bibnamefont{and}
  \bibinfo{author}{\bibfnamefont{W.}~\bibnamefont{Vogel}},
  \bibinfo{journal}{Phys. Rev. A} \textbf{\bibinfo{volume}{76}},
  \bibinfo{pages}{053835} (\bibinfo{year}{2007}).

\bibitem[{\citenamefont{Kiesel and
  Vogel}(2012{\natexlab{a}})}]{KieselVogel2011}
\bibinfo{author}{\bibfnamefont{T.}~\bibnamefont{Kiesel}} \bibnamefont{and}
  \bibinfo{author}{\bibfnamefont{W.}~\bibnamefont{Vogel}},
  \bibinfo{journal}{Phys. Rev. A} \textbf{\bibinfo{volume}{85}},
  \bibinfo{pages}{062106} (\bibinfo{year}{2012}{\natexlab{a}}).

\bibitem[{\citenamefont{Kiesel and Vogel}(2012{\natexlab{b}})}]{KieselVogel12}
\bibinfo{author}{\bibfnamefont{T.}~\bibnamefont{Kiesel}} \bibnamefont{and}
  \bibinfo{author}{\bibfnamefont{W.}~\bibnamefont{Vogel}},
  \bibinfo{journal}{Phys. Rev. A} \textbf{\bibinfo{volume}{86}},
  \bibinfo{pages}{032119} (\bibinfo{year}{2012}{\natexlab{b}}).

\bibitem[{\citenamefont{Moroder et~al.}(2010)\citenamefont{Moroder, G\"uhne,
  Beaudry, Piani, and L\"utkenhaus}}]{moroder10a}
\bibinfo{author}{\bibfnamefont{T.}~\bibnamefont{Moroder}},
  \bibinfo{author}{\bibfnamefont{O.}~\bibnamefont{G\"uhne}},
  \bibinfo{author}{\bibfnamefont{N.~J.} \bibnamefont{Beaudry}},
  \bibinfo{author}{\bibfnamefont{M.}~\bibnamefont{Piani}}, \bibnamefont{and}
  \bibinfo{author}{\bibfnamefont{N.}~\bibnamefont{L\"utkenhaus}},
  \bibinfo{journal}{Phys. Rev. A} \textbf{\bibinfo{volume}{81}},
  \bibinfo{pages}{052342} (\bibinfo{year}{2010}).

\end{thebibliography}

\bibliographystyle{apsrev}

\end{document}